\documentclass[12pt]{article}
\pdfoutput=1
\usepackage{tikz}
\usetikzlibrary{matrix,arrows,positioning,shapes,snakes,fadings,decorations.pathmorphing,
decorations.pathreplacing,automata,shadows,decorations.markings,patterns}
\usetikzlibrary{calc,shapes.callouts,shapes.arrows}
\usepackage{jheppub}
\usepackage{amsmath,amssymb,euscript,array,mathrsfs,appendix,ctable,marvosym}

\usepackage{todonotes} 
 
\newcommand \arxivlink [1]{\href{http://arxiv.org/abs/#1}{\tt arXiv:#1}}

\usepackage{titlesec}
\titleformat{\subsection}[display]{\it}{}{0.1cm}{\vspace{-1.5cm}\begin{center}\thesubsection\hspace{0.2cm}}[\end{center}\vspace{-0.5cm}]

\newcommand{\ket}[1]{|#1\rangle}

\newcommand{\EQ}[1]{\begin{equation}\begin{split} #1
\end{split}\end{equation}}

\title{Islands and Page Curves for Evaporating Black Holes in JT Gravity}
\author{Timothy J. Hollowood and S. Prem Kumar}
\affiliation{Department of Physics, Swansea University, Swansea, SA2 8PP, U.K.}
\emailAdd{t.hollowood@swansea.ac.uk, s.p.kumar@swansea.ac.uk}
\abstract{The effect of a CFT shockwave on the entanglement structure of an eternal black hole in Jackiw-Teitelboim gravity, that is in thermal equilibrium with a thermal bath, is considered. The shockwave carries energy and entropy into the black hole and heats the black hole up leading to evaporation and the eventual recovery of equilibrium.  We find an analytical description of the entire relaxational process within the semiclassical high temperature regime.
If the shockwave is inserted around the Page time then several scenarios are possible depending on the parameters. The Page time can be delayed or hastened and there can be more than one transition. The final entropy saddle has a quantum extremal surface that generically starts inside the horizon but at some later time moves outside. In general, increased shockwave energy and slow evaporation rate favour the extremal surface to be inside the horizon. The shockwave also disrupts the scrambling properties of the black hole. The same analysis is then applied to a shockwave inserted into the extremal black hole with similar conclusions.
}

\begin{document}

\maketitle

\newpage

\section{Introduction}

The black information loss paradox has inspired for over 40 years \cite{Hawking:1976ra}. Recently, however, it seems as if a step change in understanding has been achieved: it is now possible to calculate the flow of quantum information in a black hole background using only the semi-classical approximation. This new understanding grew out of holographic approaches to the gravitational entropy of the bulk theory \cite{Ryu:2006bv,Hubeny:2007xt,Faulkner:2013ana,Engelhardt:2014gca}. This theory of the ``generalized entropy" and the associated quantum extremal surfaces has now been derived from a semi-classical calculation in a black hole background via the appearance of new saddles, i.e.~instantons, known as replica wormholes \cite{Almheiri:2019qdq,Penington:2019kki}.

The replica wormhole technology, and the effective generalized entropy rules it underpins, give a new calculational window on black hole physics. In \cite{Almheiri:2019qdq} (following the earlier \cite{Almheiri:2019yqk}), a simple controllable set up was considered, consisting of an eternal (i.e.~2-sided) black hole in Jackiw-Teitelboim gravity \cite{Jackiw:1984je,Teitelboim:1983ux} in AdS$_2$ with Minkowski half spaces welded onto the boundaries, both left and right, with transparent boundary conditions. The gravity theory is coupled to a large-$c$ CFT defined over the complete spacetime, which, for simplicity, can be a free theory. The initial state of the CFT is a pure state whose left, or right, reduced state is a thermal state with the same temperature of the black hole. This ensures the whole set up is in thermal equilibrium: as the black holes evaporate Hawking modes are replaced by modes from the radiation baths at the same temperature. However, the black hole is not in entanglement equilibrium. The Hawking modes are entangled with their partners behind the horizon and this entanglement is transferred to the radiation baths as time evolves. Eventually, this entanglement entropy reaches the Bekenstein-Hawking entropy of the black hole. At this point, a new saddle, a replica wormhole, has lower---and in this case constant---entropy equal to the black hole entropy. The cross-over of entropy saddles is the semi-classical expression of the Page time of the black hole \cite{Page:1993wv}. 

The entropy transition at the Page time marks a fundamental change in the entanglement structure of the black hole. Before the transition, there is simple spatial division between the radiation and the black holes degrees-of-freedom, whereas, after the transition, an ``island" forms covering the black hole interior and, in this case, part of the exterior of the horizon. The island is secretly encoded in the radiation rather than the black hole, so the division of degrees-of-freedom is rather starkly changed. This kind of structure was guessed at some time ago and is sometimes known as the ``$A=R_B$" scenario.\footnote{The notation here refers to a Hawking mode $B$ emitted by an old black hole. It must be entangled with its partner mode $A$ behind the horizon but also with a mode in the early part of the Hawking radiation so that the final state of the radiation, after the black has evaporated, is pure. Clearly, the monogamy of entanglement does not allow this unless $A=R_B$ which implies that the inside partner mode is actually outside subtely encoded in the radiation! See \cite{Harlow:2014yka} for a detailed review of these issues.}

The beauty and simplicity of this scenario in \cite{Almheiri:2019yqk,Almheiri:2019qdq} suggests that it can provide a starting point for more detailed questions concerning the entanglement structure in a black hole background. In this work, we use the scenario to ask how the entanglement structure responds when a shockwave is created in the CFT in the radiation baths. The shockwave carries both energy and entropy. When the in-going shockwave propagates into the AdS part of the geometry and the black hole, it heats the black hole up and leads to a non-equilibrium state. Intuition suggests that the system relaxes back to equilibrium, and we confirm this. However, if the shockwave is inserted at late time around the Page time then it can change the entropy transitions in a fundamental way. The Page time can be delayed or hastened and there can be more than one transition. The structure of entropy transitions tell us the extent of the island and this determines how quickly quantum information sent into the black hole can be recovered from the radiation. Our results show that a shockwave of large energy will disrupt the scrambling of the black hole and lead to a delay in the formation of an island and the return of the quantum information carried by the shockwave. On the other hand, if the shockwave has large entropy then the formation of the island is hastened and the entanglement is returned to the radiation more quickly.

One of our main observations is that shockwave insertion into the black hole state allows us to analytically follow the complete evolution of the system whilst staying within the semiclassical regime. This is possible in a high temperature limit in which the evaporation time scale is parametrically large compared to  the inverse temperature. The limit is controlled by a saddle point approximation to Bessel functions valid for all times. 
 
The paper is set out as follows. In section \ref{s2}, we provide some of the important concepts of the entanglement structure that have emerged from recent works and establish the structure of the spacetime and details of the gravitational theory that we need. Section \ref{s3} describes the formation and properties of shockwaves in a CFT and then how the gravitational theory responds when they enter the AdS region. This will include solving in a certain limit for the dilaton of JT gravity as the shockwave propagates into the black hole. Section \ref{s4} calculates the entropy flow in the black hole plus shockwave geometry and the behaviour of the all-important quantum extremal surfaces, the boundary of the islands. In section \ref{s5}, we interpret the results of section \ref{s4} and discuss the possible entropy transitions and Page times, the scrambling times and then the interesting question of whether the quantum extremal surfaces end up inside or outside the horizon. Section \ref{s6}, applies the same analysis to a shockwave sent into an extremal black hole. Finally, in section \ref{s7} we draw some conclusions.

As this work was being completed, there appeared some related work: \cite{Gautason:2020tmk} describing the Page curve of an evaporating black hole in a related dilaton gravity model, \cite{Hashimoto:2020cas} investigating islands in Schwarzschild black holes in 4 dimensions and \cite{Bhattacharya:2020ymw} investigating islands in one dimension higher.

\section{Review: islands and the eternal black hole}\label{s2}

In this section, we briefly review the scenario that allows for the semi-classical calculation of the Page time for a black hole in JT gravity \cite{Almheiri:2019yqk} (and also \cite{Almheiri:2019qdq}). What is striking about this scenario is how it avoids the complicated back-reaction problem that would be expected for an evaporating black hole. Here there is a non-trivial transition at the Page time even in the absence of evaporation.

\subsection{The geometry}

The idea is to take the eternal black hole solution, corresponding to the extended Penrose diagram that describes a pair of black holes at a temperature $\beta^{-1}$ linked by an Einstein-Rosen bridge. The geometry is patch of AdS$_2$ with the standard metric in Poincar\'e coordinates
\EQ{
ds^2=-\frac{4dx^+dx^-}{(x^+-x^-)^2}\ .
\label{pip1}
}
In Jackiw-Teiltelboim (JT) gravity \cite{Jackiw:1984je,Teitelboim:1983ux} the metric is fixed and the non-trivial aspects of the gravitational sector involve the choice of coordinate patch and the dilaton \cite{Almheiri:2014cka,Maldacena:2016upp,Engelsoy:2016xyb,Jensen:2016pah}. The additional CFT matter fields source the dilaton rather than the metric.

\begin{figure}[ht]
\begin{center}
\begin{tikzpicture} [scale=1.1]
\filldraw[pink!20] (0,0) -- (3,3) -- (3,-3) -- (0,0);
\filldraw[yellow!20] (6,0) -- (3,3) -- (3,-3) -- (6,0);
\draw[-] (-3,-3) -- (-6,0) -- (-3,3);
\draw[-] (3,-3) -- (6,0) -- (3,3);
\draw[dashed] (-3,3) -- (3,3);
\draw[dashed] (-3,-3) -- (3,-3);
\draw[-] (-3,-3) -- (3,3);
\draw[-] (-3,3) -- (3,-3);
\draw[-] (3,-3) -- (3,3);
\draw[-] (-3,-3) -- (-3,3);
\node at (2,0) {\footnotesize AdS};
\node at (4,0) {\footnotesize bath};
\node at (-2,0) {\footnotesize AdS};
\node at (-4,0) {\footnotesize bath};
\node[right,rotate=45] at (0.3,0.6) {\footnotesize horizon $w^-=0$ };
\node[right,rotate=45] at (0.6,0.25) {\footnotesize  $x^-=\frac\beta\pi$, $y^-=\infty$};
\node[right,rotate=-45] at (-2,2.3) {\footnotesize horizon $w^+=0$};
\node[right,rotate=-45] at (0.4,-0.05) {\footnotesize $x^+=-\frac\beta\pi$, $y^+=-\infty$};
\node[right,rotate=-45] at (3.2,2.55) {\footnotesize $w^+=\infty$, $y^+=\infty$};
\node[right,rotate=45] at (3.2,-2.55) {\footnotesize  $w^-=-\infty$, $y^-=-\infty$};
\node[right,rotate=45] at (-4.9,0.9) {\footnotesize $w^-=\infty$};
\node[right,rotate=-45] at (-4.9,-0.8) {\footnotesize  $w^+=-\infty$};
%
\node[rotate=-90] at (3.2,0) {\footnotesize $w^+w^-=-1$};
\node[rotate=90] at (-2.8,0) {\footnotesize $w^+w^-=-1$};
\end{tikzpicture}
\caption{\footnotesize The coordinates of the eternal black hole pair along with their half-Minkowski space bath regions. The pink region is part of the AdS geometry outside the right black hole. The yellow region is the right bath region. The right Schwarzschild coordinates $y^\pm$ cover the pink and yellow regions. The global coordinates $w^\pm$ cover all regions on both the left and the right.}
\label{fig1} 
\end{center}
\end{figure}
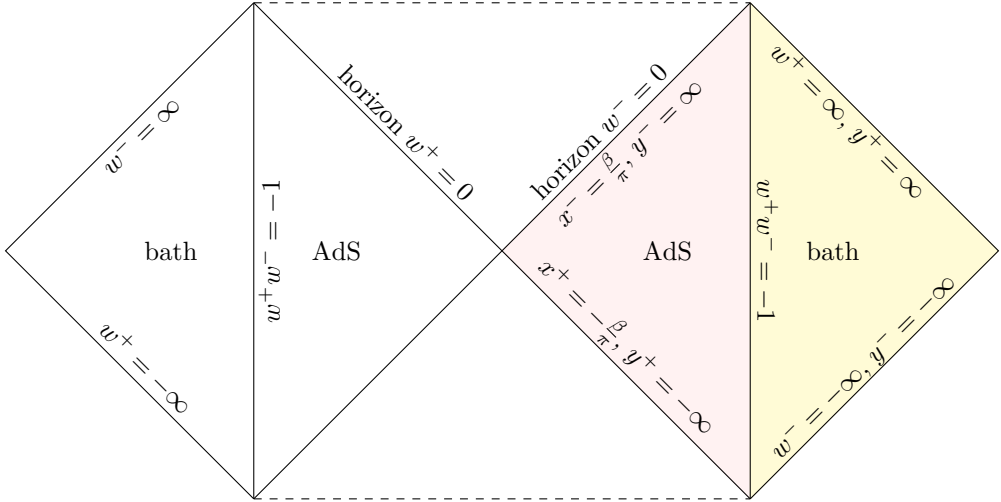

However, instead of the usual reflecting boundary conditions at spatial infinity, the geometry is extended by two half Minkowski space regions on the left and the right  \cite{Almheiri:2019yqk}. These are patched onto the AdS geometry in a smooth way. A useful set of coordinates that cover the AdS regions and also half-Minkowski regions are $w^\pm$ defined by 
\EQ{
x^\pm=\pm\frac\beta\pi\cdot\frac{w^\pm \mp1}{w^\pm\pm1}\ .
\label{lot}
}
in the AdS region, and
\EQ{
w^\pm=\pm e^{\pm 2\pi y^\pm_R/\beta}\ ,\qquad w^\pm=\mp e^{\mp 2\pi y^\pm_L/\beta}\ ,
}
in the baths. Here, $y^\pm_{L,R}$ are Minkowski coordinates with the standard metric $ds^2=-dy^+\,dy^-$, with $y^+_L-y^-_L\leq0$, and $y^+_R-y^-_R\geq0$. We will mostly consider the right black hole and bath and drop the script $R$. The coordinates are shown on the Penrose diagram in figure \ref{fig1}. The coordinates $y^\pm$ (i.e.~$y^\pm_R$), when continued into the AdS region 
\EQ{
x^\pm=\frac\beta\pi\tanh\frac{\pi y^\pm}\beta\ .
\label{wum}
}
These coordinates are ``Schwarzschild coordinates" that cover the outside of the horizon of the black hole. We write $y^\pm=t\pm r_*$, where $-\infty<r_*<\infty$ is a tortoise coordinate that covers the right AdS region for $r_*\leq0$ and the right Minkowski region for $r_*\geq0$. The time coordinate $t$ is the ``boundary time" and only the top half of the spacetime $t\geq0$  is actually relevant for the evolution of a particular initial state that is defined at $t=0$.

\subsection{The dilaton}

The dilaton is crucial to the workings of JT gravity and it plays the r\^ole that the area plays in higher dimensional black holes. In the eternal black hole background, we have
\EQ{
\phi=\phi_0+2\phi_r\frac{1-\pi^2/\beta^2 x^+x^-}{x^--x^+}=\phi_0+\frac{2\pi\phi_r}{\beta}\cdot\frac{1-w^+w^-}{1+w^+ w^-}\ .
\label{puc}
}
This solves the equations-of-motion (see e.g. \cite{Almheiri:2014cka, Almheiri:2019psf})
\EQ{
&-\frac1{(x^+-x^-)^2}\partial_{x^\pm}\big((x^+-x^-)^2\partial_{x^\pm}\phi\big)=8\pi G_NT_{x^\pm x^\pm}\ ,\\
&\partial_{x^+}\partial_{x^-}\phi+\frac2{(x^+-x^-)^2}\left(\phi-\phi_0\right)=8\pi G_NT_{x^+x^-}\ ,
\label{rit}
}
with vanishing source $T_{x^\pm x^\pm}=T_{x^+x^-}=0$. The constant $\phi_0$ sets the extremal entropy and $\phi_r$ sets a scale at which JT gravity becomes strongly coupled. The analogue of the singularity is where the dilaton vanishes, 
\EQ{
w^+w^-=\frac{2\pi\phi_r/\beta+\phi_0}{2\pi\phi_r/\beta-\phi_0}\ .
}
The Bekenstein-Hawking entropy of a black hole is determined by the value of the dilaton on the horizon $w^-=0$:
\EQ{
S_\text{BH}^{(\beta)}=\frac{\phi}{4G_N}\Big|_\text{horizon}=\frac{\phi_0+2\pi\phi_r/\beta}{4G_N}\ .
\label{ent}
}

\subsection{The quantum state}

A CFT is defined in the whole spacetime, including both the AdS and bath regions, with a large central charge $c\gg1$. It is convenient to keep things as simple as possible and choose it to be a large number $2c$ of free fermions. The fact that $c$ is large, means that the CFT modes dominate the quantum gravitational modes and the latter can be ignored in the semi-classical analysis.

On the initial value surface $t=0$ the CFT state is chosen to be the thermofield double state with respect to the left and right sides of the geometry of the same temperature as the black hole $\beta^{-1}$:
\EQ{
\ket{\psi}=\sum_ne^{-\beta E_n/2}\ket{\psi_n}_L\otimes\ket{\psi_n}_R\ .
}
Transparent boundary conditions are chosen at the boundaries between the AdS and bath regions. This ensures that CFT Hawking modes emitted by the black holes pass into their respective baths and, correspondingly, modes from the baths, at the same temperature, pass into the black holes. In this way, thermal equilibrium is maintained and there is no back-reaction on the geometry.

\subsection
{Entanglement dynamics}
 \label{sec:eebasic}
\begin{figure}[ht]
\begin{center}
\begin{tikzpicture}[scale=1]
\draw[-] (-3,3) -- (-6,0) -- (6,0) -- (3,3);
\draw[dashed] (-3,3) -- (3,3);
\draw[-] (0,0) -- (3,3);
\draw[-] (-3,3) -- (0,0);
\draw[-] (3,0) -- (3,3);
\draw[-] (-3,0) -- (-3,3);
\filldraw[black] (3,1) circle (2pt);
\filldraw[black] (-3,1) circle (2pt);
\node at (3.4,1.2) {\footnotesize $p_1$};
\node at (-3.4,1.2) {\footnotesize $p_2$};
\draw[thick] (-3,1) to[out=10,in=170] (3,1);
\draw[thick,dotted] (-6,0) to[out=20,in=-170] (-3,1);
\draw[thick,dotted] (6,0) to[out=160,in=-10] (3,1);
\node at (0,1.7) {\footnotesize $D$};
\node at (4,0.3) {\footnotesize $R$};
\node at (-4,0.3) {\footnotesize $R$};
\end{tikzpicture}
\caption{\footnotesize The calculation of the entropy of the baths region to the AdS region  involves calculating a CFT entanglement entropy for an interval $D$ across the AdS region between boundary points at a given time $t$.}
\label{fig2} 
\end{center}
\end{figure}
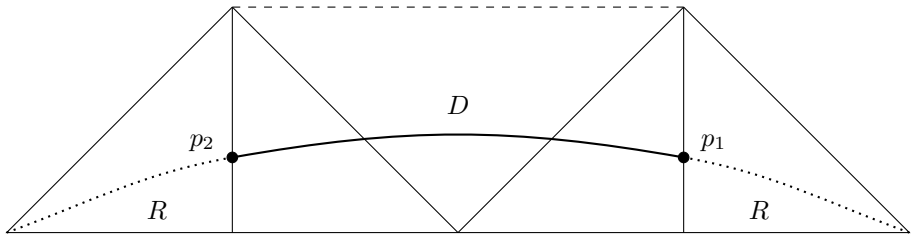

However, although thermal equilibrium is maintained, the entanglement structure of the quantum state of the CFT is not in equilibrium. As the Hawking modes are collected in the baths, the baths become more and more entangled with the black hole. This entanglement can be quantified by calculating the entanglement entropy of the baths relative to the AdS region. In the regime of parameters, this is a pure CFT calculation which is  standard \cite{Calabrese:2004eu, Calabrese:2005in}. We choose to do it at a particular time $t$ by computing the entanglement entropy of the interval across the AdS region as a subset of a complete Cauchy slice across the spacetime as shown in figure \ref{fig2} with points at
$w_1^\pm=\pm e^{\pm2\pi/\beta}$ on the right boundary and 
$w_2^\pm=\mp e^{\mp2\pi/\beta}$ on the left. Note that the boundary points are considered to be just inside the bath regions.\footnote{In \cite{Almheiri:2019yqk,Almheiri:2019qdq}, the points are taken to be at arbitrary spatial distance into the baths. Here we keep things as simple as possible and take a limit where the points move onto the boundary.} On the chosen Cauchy slice the Hilbert space factors (modulo UV issues) as ${\cal H}_R\otimes{\cal H}_D$ and the entanglement entropy of $D$, $S(D)=-\text{Tr}(\rho_D\log\rho_D)$, where $\rho_D=\text{Tr}_R|\psi\rangle\langle\psi|$. Of course, since the overall state is pure, $S(D)=S(R)$.

The calculation is straightforward once we choose an expeditious set of coordinates for which the CFT is in the vacuum state with respect to the flat metric in those coordinates. The CFT in the bath regions is in a thermal state so the stress tensor in Minkowski coordinates $y^\pm$  takes the usual thermal form $T_{y^\pm y^\pm}=\pi c/(12\beta^2)$.\footnote{Stress tensor are defined as expectation values in the semi-classical limit, normal ordered with respect to the vacuum state of the associated coordinate frame.} On the other hand, in the AdS region the stress tensor vanishes $T_{w^\pm w^\pm}=0$.\footnote{There is no anomaly for the Weyl re-scaling that takes the AdS metric to the flat metric $ds^2=-dw^+\,dw^-$.} So an appropriate choice of ``vacuum coordinates" is simply $w^\pm$ which are related to the bath coordinates by a conformal transformation
 $w^\pm=\pm e^{2\pi y^\pm/\beta}$ (on the right bath). The conformal anomaly then ensures that $T_{w^\pm w^\pm}=0$ in the bath as well. 
 The entropy is then\footnote{Here, and in the following, we ignore constant terms involving the UV cut-off of the CFT.} 
\EQ{
S(R)\equiv S(D)=\frac c6\log\left(\frac{-(w_1^+-w_2^+)(w_1^--w_2^-)}{\Omega_1\Omega_2}\right)\ .
\label{cup}
}
Here, $\Omega_{1,2}$ are conformal factors that that result from transforming the flat metric $ds^2=-dy^+\,dy^-=-\Omega^{-2}dw^+\,dw^-$ to the $w^\pm$ coordinates, so
\EQ{
\Omega^{-2}=\frac{\beta^2}{(2\pi)^2w^+w^-}\ .
}
Hence, the entropy of the radiation in the baths is
\EQ{
S(R)=\frac c3\log\Big(\frac{\beta}\pi\cosh\frac{2\pi t}\beta\Big)\ .
\label{cup2}
}
This result describes the increasing entanglement entropy of the baths as Hawking modes are collected by the bath that are entangled with their partner modes behind the horizon and in-going modes of the bath that are entangled with out-going modes. Hence,  the bath draws down entanglement from the black hole at a rate that becomes constant at late times $t\gg\beta$:
\EQ{
S\thicksim \frac {2\pi c}{3\beta} t\ .
\label{rut}
}

However, the black holes only have a finite amount of entropy to give; namely their Bekenstein-Hawking entropy $S_\text{BH}^{(\beta)}$, per black hole. At some point \eqref{rut} cannot be maintained and the entropy must top out at $2S_\text{BH}^{\beta
}$. 
This is the essence of Page's argument in this context. Note that the black holes are in thermal equilibrium and so no evaporation occurs and the final entropy should be constant.

\subsection
{Resolving the entropy paradox via replica wormholes}
 
The key insight came with the realization that in a gravitational system the von Neumann entropy is determined by a covariant variational procedure involving the ``generalized entropy" \cite{Engelhardt:2014gca} (a culmination of earlier work on holographic entropy proposals \cite{Ryu:2006bv,Hubeny:2007xt,Faulkner:2013ana}). This involves a co-dimension 2 surface $\Sigma$, the ``Quantum Extremal Surface" (QES), in terms of which
\EQ{
S_\text{gen.}(D)=\text{ext}_\Sigma\left[\frac{\text{Area}(\Sigma)}{4G_N}+S_\text{QFT}(D)\right]\ ,
}
where $D$ is the region between $\Sigma$ and the boundary of AdS and $S_\text{QFT}(D)$ is the von Neumann entropy of quantum fields on the interval $D$. One extremizes over $\Sigma$ and then if there are more than one extremum one chooses the one with the minimum entropy. In the present context of JT gravity, the QES is just a point and the r\^ole of $\text{Area}(\Sigma)$ is played by the value of the dilaton at the QES.

The generalized entropy and QES prescription was initially formulated on the basis of holography. But recent work has shown that this prescription can be derived by a semi-classical calculation of the QFT entropy across the AdS region using the replica method and allowing the extended geometry that describes the replicas to fluctuate. In the semi-classical calculation, it turns out that there are different saddles that can contribute namely, the replica wormholes \cite{Almheiri:2019qdq,Penington:2019kki}. The generalized entropy prescription turns out to be the net effect of taking into account new saddle points in the calculation of the entropy.

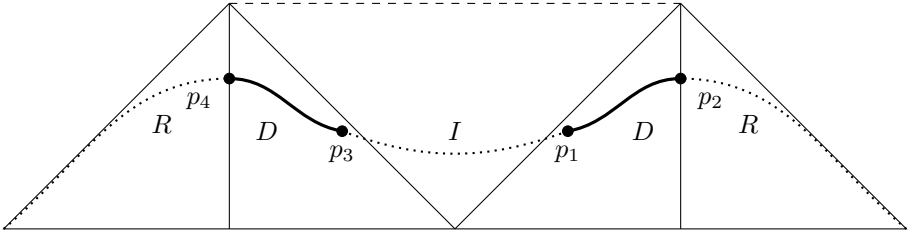
\begin{figure}[ht]
\begin{center}
\begin{tikzpicture} [scale=1]
\draw[-] (-3,3) -- (-6,0) -- (6,0) -- (3,3);
\draw[dashed] (-3,3) -- (3,3);
\draw[-] (0,0) -- (3,3);
\draw[-] (-3,3) -- (0,0);
\draw[-] (3,0) -- (3,3);
\draw[-] (-3,0) -- (-3,3);
\filldraw[black] (3,2) circle (2pt);
\filldraw[black] (-3,2) circle (2pt);
\filldraw[black] (-1.5,1.3) circle (2pt);
\filldraw[black] (1.5,1.3) circle (2pt);
\draw[very thick] (-3,2) to[out=0,in=170] (-1.5,1.3);
\draw[very thick] (3,2) to[out=-180,in=10] (1.5,1.3);
\draw[thick,dotted] (-6,0) to[out=40,in=-180] (-3,2);
\draw[thick,dotted] (-1.5,1.3) to[out=-20,in=-160] (1.5,1.3);
\draw[thick,dotted] (6,0) to[out=140,in=0] (3,2); 
\node at (3.4,1.7) {\footnotesize $p_2$};
\node at (-3.4,1.7) {\footnotesize $p_4$};
\node at (1.5,1) {\footnotesize $p_1$};
\node at (-1.5,1) {\footnotesize $p_3$};
\node at (0,1.3) {\footnotesize $I$};
\node at (2.5,1.3) {\footnotesize $D$};
\node at (3.9,1.4) {\footnotesize $R$};
\node at (-2.5,1.3) {\footnotesize $D$};
\node at (-3.9,1.4) {\footnotesize $R$};
\end{tikzpicture}
\caption{\footnotesize The generalized entropy for points $p_2$ and $p_4$ at the boundaries with QES's at points $p_1$ and $p_3$ in the bulk involves calculating the entropy of disjoint intervals as shown.}
\label{fig3} 
\end{center}
\end{figure}

In the present context, the previous calculation of the entropy of the black holes corresponds to a trivial one without a QES. However, there is a new saddle with 2 QES's, one on each side just outside the horizon, that has lower entropy at late times, shown in figure \ref{fig3}. As before, the two boundary points $p_2$ and $p_4$ are at
\EQ{
w_2^\pm=w_4^\mp=\pm e^{\pm2\pi t/\beta}
}
and there are two QES's $p_1$ and $p_3$ are located symmetrically at $w_1^\pm=w_3^\mp$. 

In order to compute the generalized entropy, we need the CFT entanglement entropy for the 2 interval configuration $D$. In general, this will depend on the cross ratio of the two points on the left and right. However, at late times to cross ratio $w_{13}w_{24}/(w_{23}w_{14})$ goes to 1.\footnote{The notation here is $w_{12}^2=(w_1^+-w_2^+)(w_1^--w_2^-)$.} With this simplification, the contribution from the left and right contributions decouple and are equal so we will concentrate on the right one and then double the result. Hence the generalized entropy is
\EQ{
S_\text{gen.}(w_1^\pm)=2\times\left\{\frac{\phi(w_1^\pm)}{4G_N}+\frac c6\log\left(\frac{-w^2_{12}}{\Omega_1\Omega_2}\right)\right\}\ .\label{genS}
}
Here, the conformal factors are
\EQ{
\Omega_1^{-2}=\frac4{(1+w_1^+w_1^-)^2}\ ,\qquad
\Omega_2^{-2}=\frac{\partial y_2^+}{\partial w_2^+}\frac{\partial y_2^-}{\partial w_2^-}=
\frac{\beta^2}{(2\pi)^2}
}
and so
\EQ{
S_\text{gen.}(w_1^\pm)=\frac{\phi_0}{2G_N}+\frac c3 F(w_1^\pm)+\frac{c}3\log \frac\beta\pi\ .
\label{luk}
} 
Here, we have defined the function $F$ to be extremized over $w_1^\pm$:
\EQ{
F(w_1^\pm)=\frac\pi{\beta k}\cdot\frac{1-w_1^+w_1^-}{1+w_1^+w_1^-}+
\log\frac{(e^{2\pi t/\beta}-w_1^+)(w_1^-+e^{-2\pi t/\beta})}{1+w_1^+w_1^-}\ .
\label{jiv}
}

The quantity 
\EQ{
k=\frac{G_Nc}{3\phi_r}\ ,
\label{sew}
}
sets the rate of evaporation of a black hole that is not in equilibrium. This must be small $k\ll1$ in order to justify the semi-classical limit.\footnote{In units of the AdS radius that has been set to 1.}  In order to simplify the analysis, we will also work in the high temperature limit for which, in addition, $\beta k\ll1$.  In this case,  the extremization is particularly simple:
\EQ{
w_1^\pm=-\frac{\beta k}{2\pi}\cdot\frac1{w_2^\mp}=\pm\frac{\beta k}{2\pi}e^{\pm2\pi t/\beta}\ .
\label{rer}
}
So the QES lies on the same constant $t$ Cauchy surface as the boundary point. The coordinate $w_1^-$ is small and negative, so the QES is just outside the horizon. On the other hand the coordinate $w_1^+$ lags behind the boundary point by an amount of time that defines the scrambling time of the black hole, i.e.~in this case
\EQ{
\Delta t_\text{s}=\frac\beta{2\pi}\log\frac{2\pi}{\beta k}\ .
\label{scr}
}
This is a realization of the Hayden-Preskill protocol \cite{Hayden:2007cs} which describes how ``diaries", i.e.~strings of qubits, thrown into an old black hole past the Page time are recovered in the Hawking radiation after a time lag, precisely $\Delta t_\text{s}$.

The critical entropy in this limit is
\EQ{
S_\text{gen.}(D)=\frac{\phi_0}{2G_N}+\frac{\pi c}{3\beta k}+\frac{c}3\log\frac\beta\pi=2S_\text{BH}^{(\beta)}\ ,
\label{cat}
}
a constant. Note that the Bekenstein-Hawking entropy includes a quantum correction from the modes of the CFT. So it is clear that the new saddle will dominant at a late time, the so-called Page time, when
\EQ{
\frac{2\pi}{3\beta} t_\text{page}\approx 2S_\text{BH}^{(\beta)}\ .
\label{fup}
}

\subsection
{The island}
 
The interpretation of the transition at the Page time is deep and far-reaching for the quantum theory of black holes. Before the Page time, the whole of the interior lies in the entanglement wedge of the pair of points $p_2$ and $p_4$ on the boundaries. This holographic concept manifests the duality between the bulk theory in the AdS part of the geometry, including the CFT, and the dual theory on the boundary of AdS. Being in the entanglement wedge of the boundary means that states and operators in a small ``code" subspace of quantum modes around the classical background are mapped to the boundary theory, at least approximately \cite{Penington:2019kki,Almheiri:2019qdq}.

\begin{figure}[ht]
\begin{center}
\begin{tikzpicture} [scale=1.1]
\filldraw[green!20] (-3,0) -- (-3,3) -- (3,3) -- (3,0) -- (-3,0);
\draw[dashed] (-3,3) -- (3,3);
\draw[-] (0,0) -- (3,3);
\draw[-] (-3,3) -- (0,0);
\draw[-] (-3,0) -- (3,0);
\draw[very thick,green] (3,0) -- (3,1.5);
\draw[very thick,green] (-3,0) -- (-3,1.5);
\draw[very thick,red] (3,1.5) -- (3,3);
\draw[very thick,red] (-3,1.5) -- (-3,3);
\filldraw[green] (3,1) circle (2pt);
\filldraw[green] (-3,1) circle (2pt);
\draw[decorate,->,decoration={snake,amplitude=0.03cm}] (1.2,0.2) -- (2.8,1.8);
\draw[decorate,->,decoration={snake,amplitude=0.03cm}] (0.2,1.2) -- (1.8,2.8);
\begin{scope}[xshift=7cm]
\filldraw[green!20]  (-3,2.5) -- (-2,1.5) -- (-3,0.5) -- (-3,2.5);
\filldraw[green!20] (3,2.5) -- (2,1.5) -- (3,0.5) -- (3,2.5);
\filldraw[red!20] (2,1.5) -- (0.5,3) -- (-0.5,3) -- (-2,1.5) -- (-0.5,0) -- (0.5,0) -- (2,1.5);
\draw[dashed] (-3,3) -- (3,3);
\draw[-] (0,0) -- (3,3);
\draw[-] (-3,3) -- (0,0);
\draw[-] (-3,0) -- (3,0);
\draw[very thick,green] (3,0) -- (3,1.5);
\draw[very thick,green] (-3,0) -- (-3,1.5);
\draw[very thick,red] (3,1.5) -- (3,3);
\draw[very thick,red] (-3,1.5) -- (-3,3);
\filldraw[red] (2,1.5) circle (2pt);
\filldraw[red] (-2,1.5) circle (2pt);
\filldraw[red] (3,2) circle (2pt);
\filldraw[red] (-3,2) circle (2pt);
\draw[decorate,->,decoration={snake,amplitude=0.03cm}] (1.2,0.2) -- (2.8,1.8);
\draw[decorate,->,decoration={snake,amplitude=0.03cm}] (0.2,1.2) -- (1.8,2.8);%
\end{scope}
\end{tikzpicture}
\caption{\footnotesize The time evolution of the entanglement structure. For early boundary times times (left), the entanglement wedge of the green boundary points consists of the entire AdS region. For late times, past the Page time, (right) the minimal entropy is captured by a configuration with 2 QES's outside the horizons. The entanglement wedges of the red boundary points are now much smaller and an island forms between the QES's (the pink region). Also shown is a Hawking mode and its partner behind the horizon. Before the Page time, the both modes are in the entanglement wedge of the boundary point whereas after the Page time the partner mode is now in the island.}
\label{fig5} 
\end{center}
\end{figure}
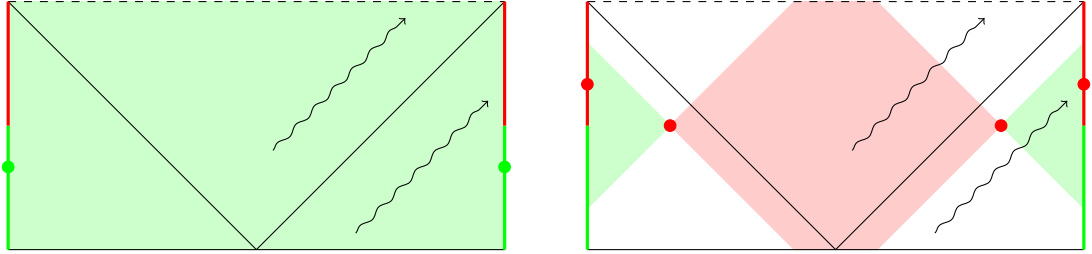

After the Page time the entanglement wedge of two boundary points $p_2$ and $p_4$ ends on the two QES's, $p_1$ and $p_3$, respectively, and so is much reduced: see figure \ref{fig5}. This leaves the wedge in between the two QES's, the so-called island $I$. Since we have calculated the entropy of the region $D$ from the boundary points to the QES's, and the overall state is pure, means that we can equate this to the entropy of the union $R\cup I$. From a holographic point of view, states and operators on the island 
are now interpreted as being ``owned" by the radiation system rather than the dual boundary theory. So the ``code" subspace of modes on $R\cup I$ is actually contained within the full Hilbert space of $R$. So the island is actually lurking in $R$ in a way that is not revealed within the semi-classical approximation. 

\section{Shockwaves}\label{s3}

\begin{figure}[ht]
\begin{center}
\begin{tikzpicture} [scale=1]
\draw[-] (-3,-3) -- (-6,0) -- (-3,3);
\draw[-] (3,-3) -- (6,0) -- (3,3);
\draw[dashed] (-3,3) -- (3,3);
\draw[dashed] (-3,-3) -- (3,-3);
\draw[-] (-3,-3) -- (3,3);
\draw[-] (-3,3) -- (3,-3);
\draw[-] (-6,0) -- (6,0);
\draw[-] (3,-3) -- (3,3);
\draw[-] (-3,-3) -- (-3,3);
\draw[decorate,very thick,->,decoration={snake,amplitude=0.03cm}] (4.5,0) -- (1.5,3);
\draw[decorate,very thick,->,decoration={snake,amplitude=0.03cm}] (4.5,0) -- (5.25,0.75);
\draw[decorate,very thick,->,decoration={snake,amplitude=0.03cm}] (-4.5,0) -- (-1.5,3);
\draw[decorate,very thick,->,decoration={snake,amplitude=0.03cm}] (-4.5,0) -- (-5.25,0.75);
\node at (2,-0.5) {\footnotesize AdS};
\node at (4,-0.5) {\footnotesize bath};
\node at (-2,-0.5) {\footnotesize AdS};
\node at (-4,-0.5) {\footnotesize bath};
\node[rotate=45] at (0.5,0.8) {\footnotesize horizon};
\node[rotate=-45] at (-0.5,0.8) {\footnotesize horizon};
\end{tikzpicture}
\caption{\footnotesize The set-up consists of the two-sided black hole in thermal equilibrium with the left and right flat space with the CFT acting as a thermal bath. Operator insertions in the baths create shockwaves that enter the AdS region and the black hole. The symmetry between the left and right is chosen to simplify the analysis.}
\label{fig6} 
\end{center}
\end{figure}
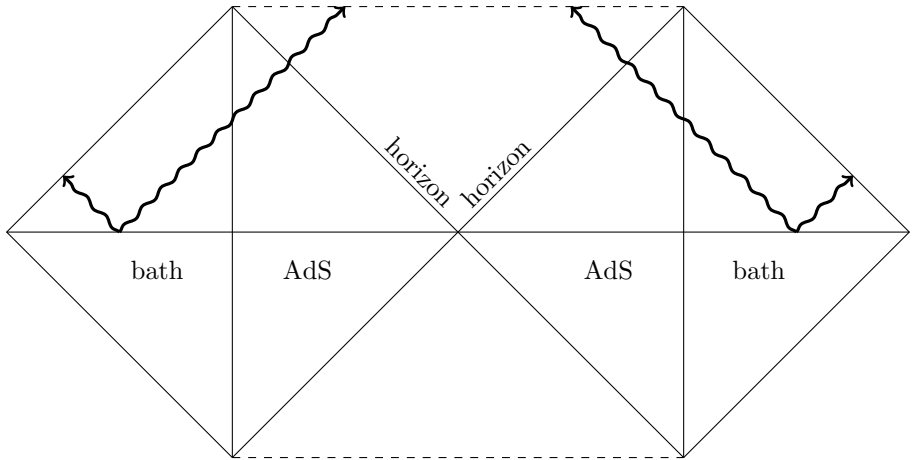

In this section, we describe how shockwaves can be generated in the bath regions that propagate along null rays into the AdS region carrying energy and entropy into the black holes. In order to simplify the analysis, we will create the shockwaves symmetrically on the left and right baths, shown in figure \ref{fig6}.

\subsection{Shockwave production}
 
 A shockwave in the AdS$_2$ region  results when a narrow pulse of energy  is sent in from  the bath. Such a pulse can be prepared by subjecting the CFT to a ``local quench" \cite{Nozaki:2014hna,Caputa:2014vaa,He:2014mwa,David:2016pzn}, i.e.~by perturbing the equilibrium state with a local operator. We can choose the operator ${\cal O}(y^+,y^-)$ to be a primary field with conformal dimension  $\Delta_{\cal O}$, so the perturbed density matrix of the right-hand side is, at $t=0$, is
 \EQ{
 \rho_{\text{Right},\varepsilon}={\cal N} {\cal O}(t_0+i\varepsilon,-t_0+i\varepsilon)\rho_\text{Right}{\cal O}^\dagger(t_0-i\varepsilon,-t_0-i\varepsilon)\ ,
}
where ${\cal N}$ is a normalization constant and $\varepsilon$ a regulator which separates the two operator insertions along the imaginary time direction.  This provides a small temporal width to the excitation thus regulating the energy in the pulse.  We will work in the limit 
 \EQ{
 \beta \gg \varepsilon ,
 }
when the resulting pulses can be approximated by left and right-moving delta-functions.\footnote{The exact universal expressions for the stress tensor expectation values, with finite width $\varepsilon$, can be found by employing standard Ward identities involving the stress tensor and two primary fields in the thermal state \cite{David:2016pzn}.} The CFT energy momentum tensor for an insertion at $y^\pm=\pm t_0$ has the form
\EQ{
T_{y^\pm y^\pm}=\frac{\pi c}{12\beta^2}+\frac{\Delta_{\cal O}}\varepsilon \delta(y^\pm\mp t_0)\ ,
\label{abc}
}
in the thermal state.\footnote{In order to be consistent with the standard normalization of the stress tensor in JT gravity, our normalization of the stress tensor differs from the usual CFT stress tensor by a factor of $\tfrac{1}{2\pi}$.}
The left-moving pulse results in a shockwave when it enters the gravitating AdS$_2$ bulk, and the energy deposited by it,
\EQ{
E_{\rm shock}=\frac{\Delta_{\cal O}}{\varepsilon}\,,
}
must be kept fixed in the limit of small $\varepsilon$. In the small width limit, the shockwave profile and strength are clearly temperature independent and so apply in the zero temperature limit as well. 

As well as having energy, the shockwaves also carry entropy due to entanglement between the left- and right-moving components. In general, we can write a chiral decomposition in the form
\EQ{
{\cal O}(y^+,y^-)=\sum_a\sqrt{p_a}\varphi_a(y^+)\bar\varphi_a(y^-)\ .
}
where the component operators are chosen to diagonalize the OPE:
\EQ{
\varphi^\dagger_a(y)\varphi_b(0)=\frac{\delta_{ab}}{y^{2\Delta_{\cal O}}}+\cdots\ .
}
The operator ${\cal O}$ creates a state with entanglement between the left and right moving sectors in the form of a Schmidt decomposition with entanglement entropy  
\EQ{
S_\text{shock}=-\sum_a p_a\log p_a\ .
\label{hut}
}
This entropy is also written as $\log d_{\cal O}$ where $d_{\cal O}$ is the quantum dimension  of the operator  $\cal O$ \cite{Nozaki:2014hna,Caputa:2014vaa,He:2014mwa}. In a typical CFT calculation, one is interested in the behaviour of the entanglement---or more generally the R\'enyi---entropy of a sub-region $A$ of space in the presence of the shockwave. As one of the components, either the left or the right, of the shockwave enters $A$ the entanglement entropy of the reduced state of $A$ jumps by $S_\text{shock}$.

For example, consider the CFT of $N$ free fermions (or $N$ copies of the Ising model),  with central charge $c=N/2$. The spin primary field $\sigma$ has $\Delta_\sigma=\frac1{16}$ and $d_\sigma=\sqrt2$.\footnote{The other primary, the energy density has $\Delta=\frac12$ and quantum dimension $d=1$.} The primary ${\cal O}=\sigma_1\cdots\sigma_N$ has scaling dimension $\Delta_{\cal O}=N/16$ and quantum dimension $d_{\cal O}=2^{N/2}=2^c$.  The shift in the entropy caused by the shockwave is then $c\log 2$.  

\subsection{Shockwaves in the AdS region}
 
We need to understand how the shockwave affects the gravitational sector as it moves from the bath to the AdS region. To our advantage, JT gravity can formulated as a theory on the boundary that boils down to a function $\tau=f(t)$ that describes the mapping between the boundary (Schwarzschild) time $t$ and the Poincar\'e time on the boundary $x^\pm=\tau$ \cite{Maldacena:2016upp,Engelsoy:2016xyb,Jensen:2016pah}. The function determines the mapping between the Poincar\'e coordinates and the Schwarzschild $x^\pm=f(y^\pm)$ of which \eqref{wum} is the example for the eternal black hole. Our analysis here has some similarity with that in \cite{Almheiri:2019psf} whose approach and notation we follow, although there are also some fundamental differences. In that reference, the shockwave is created by the coupling of the AdS spacetime to a zero temperature bath at some finite time. 

The function $f(t)$ determines the ADM energy of the bulk theory via
\EQ{
E(t)=-\frac{\phi_r}{8\pi G_N}\{f,t\}\ .
\label{ab2}
}
Here, 
\EQ{
\{f,t\}=\frac{f'''}{f'}-\frac32\left(\frac{f''}{f'}\right)^2\ ,
} 
is the Schwarzian. 

\subsection{The equilibrium state}

In the case of the eternal black hole, using \eqref{wum} at the boundary, this yields the black hole mass,
\EQ{
E_\beta=\frac{\pi\phi_r}{4\beta^2 G_N}\ .
}
The ADM mass of the spacetime satisfies an energy balance equation determined by the energy flux at the boundary, 
\EQ{
\partial_t E(t)=T_{y^+y^+}(t)-T_{y^-y^-}(t)
\label{ab1}
}
Here, $T_{y^\pm y^\pm}$ can be viewed as the normal ordered CFT stress tensor components \cite{Engelsoy:2016xyb} so that  
$T_{y^+y^+}$ is the incoming flux at the boundary, provided by the bath, and $T_{y^-y^-}$ is the outgoing flux, i.e.~the Hawking radiation.  

For the eternal black hole in thermal equilibrium with the radiation bath, the in-coming modes and out-going modes both have temperature $\beta^{-1}$ and so 
\EQ{
T_{y^\pm y^\pm}=\frac{\pi c}{12\beta^2}\ .
\label{igo}
}
In this case, the energy is constant. Note that these stress tensors are expectation values normal ordered with respect to the flat metric $ds^2=-dy^+\,dy^-$. If we transform to the Poincare coordinates in the AdS metric there is a Weyl re-scaling of the metric and a corresponding anomaly in the transformation of the stress tensor
\EQ{
T_{x^\pm x^\pm}=\Big(\frac{\partial y^\pm}{{\partial x^\pm}}\Big)^2\Big(T_{y^\pm y^\pm}+\frac c{24\pi}\{x^\pm,y^\pm\}\Big)=0\ .
\label{kka}
}
For the case of the eternal black hole, where $x^\pm$ are defined in terms of $y^\pm$ by \eqref{wum}, this means that $T_{x^\pm x^\pm}=0$ as expected.

\subsection{Effect of shockwave}
Now let us consider the effect of the shockwave that crosses the boundary at time $t_0$ with energy $E_\text{shock}$. As we have explained above, the shock is the result of a narrow pulse of energy sent in from the bath and  corresponds to modifying the in-going stress tensor $T_{y^+y^+}$ by a delta function \eqref{abc}. The effect of the shockwave is to change the map $f(t)=\beta/\pi\tanh(\pi t/\beta)$ at $t=t_0$ to some more general $f(t)$. Our task is to determine the function $f(t)$.

The in-coming modes have a stress tensor $T_{y^+y^+}$ in \eqref{abc} but what about the out-going modes for $t>t_0$? These modes will still have $T_{x^-x^-}=0$ and therefore for $x^->t_0$ we have
\EQ{
T_{y^-y^-}=-\frac{c}{24\pi}\{f(y^-),y^-\}\ .
}
Energy conservation at the boundary requires that $E(t)$ jumps by $E_\text{shock}$ at $t_0$ and then for $t>t_0$ satisfies 
\EQ{
\partial_tE(t) =\frac{\pi c}{12\beta^2}+\frac c{24\pi}\{f(t),t\}\ .
\label{ab1}
}
The equation is solved with the boundary condition $E(t_0)=E_\beta+E_\text{shock}$. It is useful to parametrize the shockwave energy in terms of a new, higher temperature $\tilde\beta^{-1}$, so that
\EQ{
E_\text{shock}=E_{\tilde\beta}-E_\beta\ .
\label{dun}
}
Intuitively, the shockwave raises the black hole temperature to $\tilde\beta^{-1} > \beta^{-1}$, which  is then expected to evaporate back to the original thermal state at temperature $\beta^{-1}$. We will shortly prove this intuition is correct. 
 
Combining eq.\eqref{ab2} for the energy flux with the ADM energy \eqref{ab1} we obtain,
\EQ{
\partial_tE(t)=\frac{\pi c}{12\beta^2}-k E(t)\ ,
\label{bor}
}
where $k$ was defined in \eqref{sew}. Solving, for $t>t_0$,
\EQ{
E(t)=\frac{\pi\phi_r}{4G_N}\left(\beta^{-2}+(\tilde\beta^{-2}-\beta^{-2})\,e^{-k(t-t_0)}\right)\ .
\label{tup}
}
Therefore the black hole settles back to the thermal state at temperature $\beta^{-1}$ beyond  a time scale $k^{-1}$ after the injection of the shock. 

\subsection{The exact solution}

In order to determine the complete background and dilaton, we need to solve for the function $f(t)$. The exponential falloff of the ADM energy implies that the key function $f(t)$ solves the third order differential equation 
\EQ{
\{f(t),t\}=-2\pi^2\left(\beta^{-2}+(\tilde\beta^{-2}-\beta^{-2})e^{-k(t-t_0)}\right)\,,\qquad t> t_0\,.
\label{lop}
} 
Importantly, solutions to this differential equation are only determined up to a M\"obius transformation 
\EQ{
f\to\frac{Af+C}{Cf+D}\ ,
\label{xix}
}
whose freedom corresponds to the three integration constants of the third order differential equation. The M\"obius transformation is determined by requiring that $f(t)$ is continuous up to its second derivative across $t=t_0$. Given that $f(t)=\beta/\pi \tanh(\pi t/\beta)$, for $t<t_0$, gives the conditions 
\EQ{
f(t_0)&=\frac\beta\pi\tanh\frac{\pi t_0}\beta\ ,\\ f'(t_0)&=\text{sech}^2\frac{\pi t_0}\beta\ ,\\f''(t_0)&=-\frac{2\pi}\beta \sinh\frac{\pi t_0}\beta\,\,\text{sech}^3\frac{\pi t_0}\beta\ ,
\label{pop}
}
which, given a particular solution of \eqref{lop}, fixes the freedom in \eqref{xix}. 

A particular integral of \eqref{lop} is expressed in terms of modified Bessel functions of the 1st and the 2nd kind:
\EQ{
\hat f=\alpha\frac{K_\nu(\nu z)}
{I_\nu(\nu z)}\ ,\qquad \nu=\frac{2\pi}{\beta k}\ ,\quad z=\sqrt{\frac{E_\text{shock}}{E_\beta}}\,e^{-k(t-t_0)/2}\ .
\label{wup}
}
The constant $\alpha$ is a convenient normalization that we fix below by requiring $\hat f(t_0)=e^{2\pi t_0/\tilde\beta}$.
The example in \cite{Almheiri:2019psf} corresponds to the particular case $\beta\to\infty$, so $\nu=0$ with $\nu z$ fixed. We will discuss this case, separately in section \ref{s6} where it corresponds to a shockwave incident on an extremal black hole.

The exact formulae for the integration constants in the M\"obius transformation are presented in  appendix \ref{sec:appadia}, and we find,
\EQ{
f(t) =\frac{\beta}{\pi}\cdot\frac{K_\nu(\nu z_0)\big(\hat f (t)/\hat f (t_0)-1\big)+z_0\tanh\tfrac{\pi t_0}{\beta}\big(\hat f(t)I_\nu^\prime(\nu z_0)/\alpha-K_\nu^\prime(\nu z_0)\big)}{K_\nu(\nu z_0)\big(\hat f (t)/\hat f (t_0)-1\big)\tanh\tfrac{\pi t_0}{\beta}+z_0\big(\hat f(t)I_\nu^\prime(\nu z_0)/\alpha-K_\nu^\prime(\nu z_0)\big)}\ .
\label{exactf}
}
In the late time limit, $t\gg k^{-1}$, when $z$ is small, the asymptotics of Bessel functions imply that
\EQ{
\hat f(t)\left.\right|_{z\ll 1} \propto e^{2\pi t/\beta}\,, 
}
which is what we expect for a black hole that has relaxed back towards its original thermal state.

\subsection{High temperature limit}

For the semiclassical approximation to apply, we must keep $k\ll 1$ (defined in \eqref{sew}) since it controls matter loops and also sets the black hole evaporation time scale.\footnote{The parameter $k$ is the effective coupling between the boundary degree of freedom in AdS$_2$ and the CFT  \cite{Engelsoy:2016xyb,Almheiri:2019psf}. The ADM energy and associated flux equation \eqref{bor} assumes the small $k$ semi-classical limit.} On the other hand, the combination $\beta k$ can be arbitrary. However, we will work in the regime where $\beta k\ll1$, so the index $\nu$ for the Bessel functions is large. This is justified when the temperature is sufficiently high and so we can call it the ``high temperature limit". Working in this regime simplifies the analysis as we can employ the saddle-point evaluation of the integral representation of the modified Bessel function to derive the following approximate form (see appendix \ref{sec:appadia}). 

With a suitable fixing of the normalization $\alpha$ in \eqref{wup}, this gives the small $\beta k$, i.e.~$\nu\gg1$, approximate form
\EQ{
\hat f(t) =e^{2\nu({\cal S}(t)-{\cal S}(t_0))+2\pi t_0/\tilde\beta}\ ,\qquad
{\cal S}(t)\equiv-\sqrt{1+z^2}+\tanh^{-1}\frac1{\sqrt{1+z^2}}\ ,
\label{jip}
}
where $z$ is given as a function of $t$ in \eqref{wup}. The normalization has been fixed by making the convenient choice that $\hat f (t_0)=e^{2\pi t_0/\tilde\beta}$. The behaviour of $\hat f$ in the neighbourhood of $t_0$ is then
\EQ{
\log\hat f(t)=\frac{2\pi t}{\tilde\beta}-\frac{\pi k(\beta^2-\tilde\beta^2)}{2\beta^2\tilde\beta}(t-t_0)^2+{\mathscr O}((t-t_0)^3)\,.
\label{bid}
}
The exponential dependence on time immediately after the shockwave injection is consistent with a black hole at a new, higher temperature $\tilde\beta^{-1}$.  In the limit of small $\beta k$, this lasts for a time scale $\sim {\mathscr O}\left( k^{-1}\right)$.

In the high temperature limit, our exact solution \eqref{exactf} for the map $f(t)$, obtained by the continuity conditions at $t=t_0$, yields
\EQ{
f(t)=\frac{\beta}\pi\cdot\frac{\beta \tanh\frac{\pi t_0}\beta+\tilde\beta\tanh\left[\nu\left({\cal S}(t)-{\cal S}(t_0)\right)\right] }{\beta+\tilde\beta \tanh\left[\nu\left({\cal S}(t)-{\cal S}(t_0)\right)\right]\tanh\frac{\pi t_0}\beta}\ .
\label{ab4}
} 
This result satisfies a simple check. In the limit that $\tilde \beta$ approaches $\beta$,
\EQ
{ \lim_{\tilde \beta\to \beta} \nu\left({\cal S}(t)-{\cal S}(t_0)\right)=\frac{\pi}{\beta}(t-t_0)\,, 
}
and we recover $f(t)=\beta/\pi \tanh\left(\pi t/\beta\right)$ which is the equilibrium result. 
 
In the late time limit, when $t\gg k^{-1}$ we find,
\EQ{
\log\hat f(t)= \frac{2\pi}{\beta} (t+\kappa)+{\mathscr O}(e^{-k(t-t_0)})\ ,
\label{bom}
}
where the constant
\EQ{
\kappa=-\frac{2}{k}\Big(1+\log\frac{\sqrt{\beta^2-\tilde\beta^2}}{2\tilde\beta}-\frac{k\beta t_0}{2\tilde\beta}-\frac\beta{\tilde\beta}+\tanh^{-1}\frac{\tilde\beta}\beta\Big)\ .
}
The behaviour \eqref{bom} is exactly as we would expect for a black hole of the original temperature $\beta^{-1}$. Therefore, our saddle point expressions correctly capture the relaxation of the black hole to the original thermal state after (slow) evaporation.

After the shockwave is sent in, the horizon of the black hole shifts outwards from $x^-=\beta/\pi$ to $x^-=f(\infty)$. Within the high temperature limit, this is
\EQ{
x^-_\text{hor.}=f(\infty)=\frac\beta\pi\cdot\frac{\tilde\beta+\beta\tanh\frac{\pi t_0}\beta}{\beta+\tilde\beta\tanh\frac{\pi t_0}\beta}\, <\, \frac{\beta}{\pi}\,.
\label{hit}
}
Thus the horizon shifts even though the black hole returns to equilibrium.

The function $f(t)$ which determines the relation between boundary  time $t$ and Poincar\'e  time, naturally extends into the AdS$_2$ bulk. It is now natural to define new coordinates behind the shockwave,
\EQ{
\tilde x^\pm=\frac{\tilde\beta}\pi\cdot\frac{\hat f(y^\pm)-1}{\hat f(y^\pm)+1}=\frac{\tilde\beta}\pi\tanh\left[\nu({\cal S}(y^\pm)-{\cal S}(t_0))+\frac{\pi t_0}{\tilde\beta}\right]\,.
\label{cut}
}
Immediately after the shockwave, it follows from eq.\eqref{bid} that $\tilde x^\pm=\tilde\beta/\pi\tanh(\pi y^\pm/\tilde\beta)$, matching on to the  eternal black hole patch \eqref{wum} but with a higher temperature $\tilde\beta^{-1}$.  Furthermore, \eqref{cut}  reveals that the subsequent relaxation of the black hole is characterized by an effective temperature $\beta_{\rm eff}^{-1}$,
\EQ
{
\tilde x^\pm=\frac{\tilde \beta}{\pi}\tanh\left[\pi\int_{t_0}^t\frac{dt^\prime}{\beta_{\rm eff}(t^\prime)}\,+\frac{\pi t_0}{\beta}\right]\,.
}
The effective temperature decreases monotonically from $\tilde\beta^{-1}$ towards the original unperturbed value and is given by 
\EQ
{
\frac{1}{\beta_{\rm eff}(t)}=\frac{1}{\beta}\sqrt{1+ \left(E_{\rm shock}/{E_{\beta}}\right)\, e^{-k(t-t_0)}}\,.\label{Teff}
}
In the calculations we present below, the associated coordinates
\EQ{
\tilde w^\pm=\pm\hat f(y^\pm)^{\pm1}\,,\label{tildew}
}
prove to be useful. In terms of the new coordinates, the horizon is always at $\tilde x^-=\tilde\beta/\pi$ or  $\tilde w^-=0$ with $\tilde w^-\gtrless0$ being inside/outside. 

\subsection{Stress tensor}
 
In order to solve for back-reaction on the dilaton after the shockwave enters, we need to know the stress tensor. For the out-going modes, there is no change and in Poincar\'e frame $T_{x^-x^-}=0$ always. For the in-going modes, the stress tensor is \eqref{igo} which we can transform to the Poincar\'e frame using $x^+=f(y^+)$ and \eqref{kka} 
\EQ{
T_{x^+x^+}&=\Big(\frac{\partial y^+}{\partial x^+}\Big)^2\left(\frac{\pi c}{12\beta^2}\,+\,E_{\rm shock}\delta(y^+-t_0)\right)\,-\,\frac c{24\pi}\{f^{-1}(x^+),x^+\}\ .
\label{hut2}
}
Using \eqref{lop} we then find the explicit form,
\EQ{
T_{x^+x^+}=\frac{\beta^2-\tilde\beta^2}{\beta^2\tilde\beta^2}\Big(\frac{\pi\phi_r}{4G_N}\cosh^2\frac{\pi t_0}\beta\delta(x^+-x_0)-\frac{\pi c\, e^{-k(y^+-t_0)}}{12 f'(y^+)^2}\theta(x^+-x_0)\Big)\ ,
\label{hut}
}
which determines the discontinuity in the derivative of the dilaton through \eqref{rit}. Here $x_0=\beta/\pi\tanh(\pi t_0/\beta)$. 

There is a useful way to  rewrite \eqref{hut2}  behind the shockwave, when $x^+>x_0$ (or $y^+>t_0$). Since
$\{f^{-1}(x^+),x^+\}=-(f'(y^+))^{-2}\{f(y^+),y^+\}$, for $x^+=f(y^+)$, we have, as a function of $y^+$
\EQ{
T_{x^+x^+}=\frac1{f'(y^+)^2}\Big(\frac{\pi c}{12\beta^2}+\frac c{24\pi}\{f(y^+),y^+\}\Big)\ .
}
But from \eqref{ab1}, this is
\EQ{
T_{x^+x^+}=\frac1{f'(y^+)^2}\partial_{y^+}E(y^+)=-\frac{\phi_r}{8\pi G_N}\cdot\frac1{f'(y^+)^2}\partial_{y^+}\{f(y^+),y^+\}\ .
}
Converting the $y^+$ to $x^+$ derivatives in the Schwarzian, and assuming $y^+=y^+(x^+)$, this yields
\EQ{
T_{x^+x^+}=-\frac{\phi_r}{8\pi G_N}\partial_{x^+}^3f'(y^+)\ .
\label{dii}
}
This general result will prove useful when we solve for the dilaton. 

\subsection{Vacuum coordinates}
 
The strategy for calculating the von Neumann entropy of the bath will be to relate the relevant  CFT correlators to corresponding vacuum correlators by an apporpriate conformal transformation.
 Therefore the key to this calculation  is to find a coordinate frame for which the CFT is in the vacuum state (a summary of various coordinate systems employed is provided in appendix \ref{appcoords}). Importantly, these frames are only defined up to a M\"obius transformation which can be chosen for convenience. 

The in-going modes in the original Schwarzschild frame $y^+$ are always in a state of temperature $\beta^{-1}$. These modes  can  therefore be mapped to the vacuum CFT state by the exponential map,  choosing $e^{2\pi y^+/\beta}$ as a vacuum coordinate. For $y^+<t_0$, this is
\EQ{
w^+=e^{2\pi y^+/\beta}\ .
}
On the other hand, the out-going modes in the Poincar\'e frame $x^-$ are in the vacuum state, so we can take $w^-$ or, indeed, $\tilde w^-$, both related to $x^-$ by a M\"obius transformation. For $y^-<t_0$, we have 
\EQ{
w^-=-e^{-2\pi y^-/\beta}\ ,
} 
showing that before the shockwave the outgoing Hawking modes have temperature $\beta^{-1}$. Behind the shockwave, for 
$y^->t_0$, the $w^-$ coordinate is related to $y^-$ by\footnote{Note that in the absence of a shockwave, $f(y^-)=\beta/\pi\tanh(\pi y^-/\beta)$ and $\eta(y^-)=y^-$.
}
\EQ{
w^-\equiv-e^{-2\pi\eta(y^-)/\beta}=\frac{\pi f(y^-)-\beta}{\pi f(y^-)+\beta}\ .
\label{wit}
}
Within the high temperature approximation $\beta k\ll1$, and in terms of $\hat f(t)$, we have
\EQ{
e^{-2\pi\eta(t)/\beta}=e^{-2\pi t_0/\beta}\frac{(\beta-\tilde\beta)\hat f(t)+(\beta+\tilde\beta)e^{2\pi t_0/\tilde\beta}}
{(\beta+\tilde\beta)\hat f(t)+(\beta-\tilde\beta)e^{2\pi t_0/\tilde\beta}}\ .
\label{yaa}
}
The behaviour of $\eta(t)$ is important in our analysis. Recalling that in the high temperature limit $\hat f(t)$ rises exponentially as $\sim e^{2\pi t/\tilde\beta}$ at early times, $\eta(t)$ increases linearly from $t=t_0$ but then at a later time,
\EQ{
t\thicksim t_0+\frac{\tilde\beta}{2\pi}\log\frac{4\beta\tilde\beta}{\beta^2-\tilde\beta^2}\ ,
\label{tur}
}
saturates at the value
\EQ{
\eta_0=t_0+\frac{\beta}{2\pi}\log\frac{\beta+\tilde\beta}{\beta-\tilde\beta}\ .
\label{his}
}
For the out-going modes, we can also use $\tilde w^-$ as the vacuum coordinate, where
\EQ{
\tilde w^-=-\frac1{\hat f(y^-)}
\label{ch1}
}
which is a useful coordinate behind the shockwave. This is natural in the high temperature limit, wherein,
$\tilde w^- = -1/\hat f(y^-) \approx - e^{- 2\nu({\cal S}(y^-)-{\cal S}(t_0))-2\pi t_0/\tilde\beta}$.

\subsection{Dilaton}
 
In the semi-classical limit, the dilaton is sourced by the expectation value of the stress tensor of the CFT \eqref{rit}.
As we cross the shockwave, $T_{x^+x^+}$ has a delta function and then a non-vanishing contribution. This would seem to make the problem of solving for the dilaton a complicated problem. Fortunately with a non-vanishing $T_{x^+x^+}$ but with $T_{x^-x^-}=T_{x^+x^-}=0$, there is a simple expression for the dilaton in term of the key function $f(t)$ \cite{Moitra:2019xoj}. In order to find it, we note that the $--$ and $+-$ components of the equations for the dilaton \eqref{rit}, have a general solution of the form
\EQ{
\phi=\phi_0+2\phi_r\Big(\tfrac12\partial_{x^+}h(x^+)+\frac{h(x^+)}{x^--x^+}\Big)\ .
}
Then the $++$ equation gives
\EQ{
T_{x^+x^+}=-\frac1{8\pi G_N}\partial_{x^+}^3h(x^+)\ .
}
Now we compare with the expression for this stress tensor component in \eqref{dii}. Clearly we have perfect agreement if we identify\footnote{Strictly, the identification is modulo three integration constants $h(x^+)=f'(y^+)+ a_0 + a_1 x^+ + a_2 (x^+)^2$ which vanish by requiring matching with the new thermal state immediately after $t=t_0$.}
\EQ{
h(x^+)=f'(y^+)\ ,
}
where $x^+=f(y^+)$. So, implicitly, in solving the equation at the boundary for energy balance, we have implicitly solved for the dilaton. Importantly,  we can avoid having to use Green function methods and memory integrals. Note that the delta-function singularity in $T_{x^+ x^+}$ and its strength \eqref{hut} follows automatically from the discontinuity in $f'''(y^+)$, implied by the equation for the Schwarzian \eqref{lop}.

We can now write the dilaton in mixed coordinates $(y^+,x^-)$ explicitly as
\EQ{
\phi=\phi_0+2\phi_r\left(\frac{f''(y^+)}{2f'(y^+)}+\frac{f'(y^+)}{x^--f(y^+)}\right)\ .
\label{les}
}
In front of the shockwave, $f(t)=\beta/\pi\tanh(\pi t/\beta)$, and one finds \eqref{puc}.

Behind the shockwave, we find it more useful to use the mixed coordinates $(y^+,\tilde w^-)$ where $\tilde w^+=\hat f(y^+)$. After a M\"obius transfromation trading $f(y^\pm)$ for $\hat f (y^\pm)$, the dilaton is then
\EQ{
\phi=\phi_0+2\phi_r\left(\frac{\hat f''(y^+)}{2\hat f'(y^+)}-\frac{\tilde w^-\hat f'(y^+)}{1+\tilde w^-\hat f(y^+)}\right)\ .
\label{los}
}
Just after the shockwave, in the high temperature limit, we have $\hat f(t)=e^{2\pi t/\tilde\beta}$ ($t\approx t_0$ and $\beta k\ll 1$), in which case the dilaton takes the form
\EQ{
\phi=\phi_0+\frac{2\pi\phi_r}{\tilde\beta}\cdot\frac{1-\tilde w^+\tilde w^-}{1+\tilde w^+\tilde w^-}\ ,
\label{puc2}
}
exactly what one would expect for a black hole of a new temperature $\tilde\beta^{-1}$.

\subsection{Entropy of the evaporating black hole}

The dilaton determines the entropy of the black hole \eqref{ent}. After the passage of the shockwave, the horizon is at $\tilde x^-=\tilde\beta/\pi$ (i.e.~$\tilde w^-=0$). Inserting this into \eqref{los} gives a remarkably simple expression for the entropy as a function of the boundary time of an in-going null ray with coordinate $y^+$:
\EQ{
S_\text{BH}(y^+)=\frac1{4G_N}\left(\phi_0+\phi_r\frac{\hat f''(y^+)}{\hat f'(y^+)}\right)\ .
\label{due}
}
This gives
\EQ{
S_\text{BH}(y^+)=\frac1{4G_N}\left(\phi_0+\frac{k\phi_r}{z}\cdot\frac{(\nu z^2+2(\nu-1))I_{\nu-1}(\nu z)-\nu z I_{\nu-2}(\nu z)}{I_\nu(\nu z)}\right)\ ,
}
where $\nu$ and $z$ are defined in \eqref{wup} with $t$ replaced by the null coordinate $y^+$. 
Using the approximation for $\hat f$ in \eqref{jip}, valid for small $\beta k$, gives the explicit expression
\EQ{
S_\text{BH}(y^+)\Big|_{\beta k\ll 1}\,=\frac1{4G_N}\left(\phi_0+\frac{2\pi\phi_r}{\beta_{\rm eff}(y^+)}
\right)\ ,
}
written naturally in terms of the effective temperature \eqref{Teff}, and  exhibiting monotonic decrease from $S_\text{BH}^{(\tilde\beta)}$ to $S_\text{BH}^{(\beta)}$. This proves that the geometry settles down to a black hole at the original temperature.

\section{Entropy saddles}\label{s4}

In this section, we consider the effect of the shockwave on the entropy of the radiation. Let us emphasize the approximations we are making. These are done to avoid numerical solutions and so make the interpretation of the results more transparent:

\noindent (i) We work in the limit $\beta k\ll1$. This simplifies the equations that determine the position of the QES and also allow us to use the approximate form for the map $\hat f(t)$ in \eqref{jip}. 

\noindent (ii) As in \cite{Almheiri:2019yqk,Almheiri:2019qdq}, since we are interested in late-time phenomena, around or after the Page time of the original black hole, we shall ignore the cross terms in the entanglement entropy that link the left and the right systems. In particular, this follows from the fact that relevant time scales, including $t_0$, scale like $k^{-1}$. We will discuss the validity of this procedure ex post facto in section \ref{s4.4}.

\subsection{No Islands}
 
The no-island entropy reviewed in section \ref{sec:eebasic} is valid before the boundary point crosses the in-coming shockwave at $t=t_0$. The calculation was performed using the vacuum coordinate frame $w^\pm$. As the boundary point crosses the shockwave we can still use $w^\pm$ as vacuum coordinates and so in terms of these coordinates the result is not changed. However, the mapping of the coordinate $w^-$ to the boundary time changes:
\EQ{
w^+=e^{2\pi t/\beta}\ ,\qquad w^-=-e^{-2\pi\eta(t)/\beta}\ ,
\label{cik}
}
where $\eta(t)$ was defined in \eqref{yaa}. This changes the conformal factor of the boundary point to
\EQ{
\Omega^{-2}=\frac{\partial y^+}{\partial w^+}\frac{\partial y^-}{\partial w^-}\Big|_{y^\pm=t}=
\frac{\beta^2}{(2\pi)^2}e^{-2\pi(t-\eta(t))/\beta-\log\eta'(t)}\ .
\label{xee}
}
Plugging these into \eqref{cup} gives 
\EQ{
S_\text{no island}&=\frac c3\log\left(e^{2\pi t/\beta}+e^{-2\pi\eta(t)/\beta}\right)-\frac{\pi c}{3\beta}(t-\eta(t))\\ &-\frac c6\log\eta'(t)+
\frac c3\log\frac\beta{2\pi}+2S_\text{shock}\ .
}
From $t=t_0$, $\eta(t)$ grows linearly at very early times, $\eta(t)\simeq t$, and the entropy matches that of the no-island contribution before the shockwave \eqref{rut} but with a shift by the shockwave entropy \eqref{hut}. This is to be expected, as the shockwave enters the AdS region the entanglement entropy jumps because the radiation bath is the purifier of shockwave.

After the time \eqref{tur}, $\eta(t)$ saturates to the constant value $\eta_0$ \eqref{his}. However, the function $\hat f(t)$ continues to grow \eqref{bid} as $\hat f\sim \exp(2\pi t/\tilde\beta)$, characteristic of the higher effective temperature after the injection of the shockwave. This growth continues for a parametrically long  timescale $t<{\mathscr O}(k^{-1})$,  well after $\eta(t)$ has saturated. During this phase the entropy continues to increase linearly
\EQ{
S_\text{no island}\simeq\frac{\pi c}{3\tilde\beta}(t-t_0)+\frac{\pi c}{3\beta}(t+t_0)\,+\frac c3\log\frac{\beta+\tilde\beta}{4\pi}+2S_\text{shock}\ .
\label{xie}
}
At later times $t\gg k^{-1}$, the black hole relaxes towards its original temperature and $\hat f(t)\,\sim\, \exp(2\pi(t+\kappa)/\beta)$ as per eq.\eqref{bom}, and the rate  of growth of the no-island entanglement entropy (entanglement velocity) changes:
\EQ{
S_\text{no island}\,\simeq\,\frac{2\pi c}{3\beta}t+\frac{\pi c}{3\beta}(t_0+\kappa)\,-\frac{\pi ct_0}{3\tilde\beta}+\frac c3\log\frac{\beta+\tilde\beta}{4\pi}+
\frac c6\log\frac\beta{\tilde\beta}+2S_\text{shock}\ .
\label{xif}
}

\subsection{Island with QES in front of shockwave}
 
\begin{figure}[ht]
\begin{center}
\begin{tikzpicture} [scale=0.65]
\filldraw[red!10] (-2.5,4) -- (-4.5,6) -- (-10,6) -- (-10,-3.5) -- (-2.5,4);
\filldraw[blue!10] (-8,-0.5) -- (-10,1.5) -- (-10,-2.5) -- (-8,-0.5);
\filldraw[green!10] (-9,-2) -- (-10,-1) -- (-10,-3) -- (-9,-2);
\draw[very thick,green] (0,-3.5) -- (0,0);
\draw[very thick,blue] (0,0) -- (0,4);
\draw[very thick,red] (0,4) -- (0,6);
\draw[-] (3,3) -- (0,6);
\draw[-] (-3,3) -- (0,6);
\draw[-] (-4,4) -- (-10,-2);
%
%
\filldraw[green] (0,-1) circle (4pt);
\filldraw[blue] (0,2) circle (4pt);
\filldraw[red] (0,4.5) circle (4pt);
\filldraw[green] (-9,-2) circle (4pt);
\filldraw[blue] (-8.2,-0.3) circle (4pt);
\filldraw[red] (-2.5,4) circle (4pt);
\draw[blue,dashed] (-8.5,0) -- (-7,-1.5);
\draw[decorate,very thick,->,decoration={snake,amplitude=0.03cm}] (2,-2) -- (-5.8,5.8);
\node at (2,0) {\footnotesize bath};
\node at (-3,0) {\footnotesize AdS};
\draw[very thick,blue,dashed] (-8.85,-1.5) to[out=70,in=-125] (-8.2,-0.3);
\draw[very thick,green,dashed] (-9.3,-3.5) to[out=85,in=-110] (-8.85,-1.5);
\draw[very thick,red,dashed] (-3.4,3.6) to[out=20,in=-135] (0,6);
\node at (-7.5,2) (a1) {\footnotesize $w^-=0$};
\node at (-2.5,5.5) (a2) {\footnotesize $\tilde w^-=0$};
\draw[->] (a1) -- (-7.1,1);
\draw[->] (a2) -- (-2.1,4);
\node at (-6.5,-2) {\footnotesize $w^+_\text{frozen}$};
\end{tikzpicture}
\caption{\footnotesize A schematic plot of the  3 possible island saddle points corresponding to par	ticular boundary points with the QES shown as blobs and the islands shown as the shaded regions. The boundary regions are coloured according to which type of QES has the minimum entropy and the motion of the QES's are shown as the dotted lines. The green blob lies on the same Schwarzschild Cauchy slice as its boundary point, whereas the blue and the red lag behind. The null coordinate $w^+$ of the blue QES becomes frozen at roughly a scrambling time before $t_0$. For large enough shockwave energy, the red QES lies behind the shifted horizon as shown here.}
\label{fig7} 
\end{center}
\end{figure}
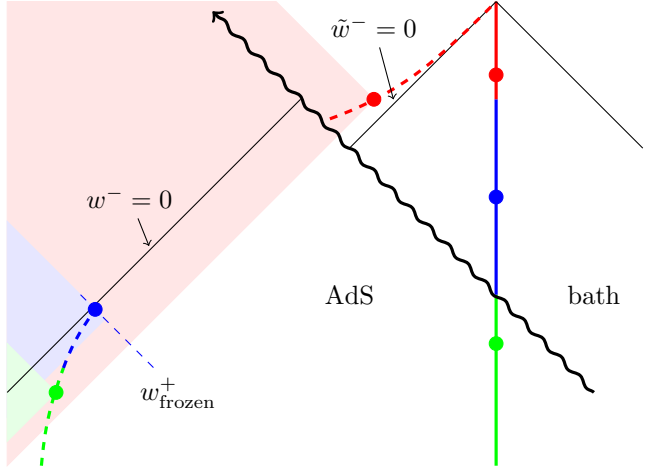

Now we consider the entropy contributions from configurations {\em with} an island. These are illustrated in figure \ref{fig7}. Previously we reviewed the calculation of the entropy with an island leading to \eqref{cat}. This corresponds to the green boundary region and QES in figure \ref{fig7}. This result changes when the boundary point crosses the in-coming shockwave and enters the blue region on the boundary.

There are two distinct situations to consider,  the first in which the quantum extremal surface resides in front of the shockwave in the unperturbed portion of the geometry, and a second scenario wherein it lies behind the shockwave.  Which of these two kinds of configurations appears depends on the time elapsed after injection of the shockwave.

We can still use vacuum coordinates $w^\pm$ but now the mapping  $w^-_2$ for the boundary point to the boundary time changes as in \eqref{cik}
\EQ{
w^+_2=e^{2\pi t/\beta}\ ,\qquad w^-_2=-e^{-2\pi\eta(t)/\beta}\ .
\label{cyr}
}
The conformal factor of the boundary point also  changes as in \eqref{xee}.

Taking the QES to lie in front of the shockwave, the expression for the entropy is given by \eqref{genS} but with the modified mapping \eqref{cyr} and conformal factor for the boundary point, including a jump from the shockwave entropy,
\EQ{
S_\text{gen.}(w_1^\pm)=\frac{\phi_0}{2G_N}+\frac c3 F(w_1^\pm)+\frac c3\log\frac{2}{\Omega_2}+2S_\text{shock}\ .
} 
The function $F$, to be extremized over $w_1^\pm$, is 
\EQ{
F(w_1^\pm)=\frac\pi{\beta k}\cdot\frac{1-w_1^+w_1^-}{1+w_1^+w_1^-}+
\log\frac{(-w_1^++e^{2\pi t/\beta})(w_1^-+e^{-2\pi\eta(t)/\beta})}{1+w_1^+w_1^-}\ ,
}
which is identical to \eqref{jiv} apart from the change of mapping of the boundary coordinate $w^-_2$. This is because the QES is in front of the shockwave and the form of the dilaton is unchanged.
 In the high temperature limit $\beta k \ll 1$, the extremization over $w_1^\pm$ again gives $w_1^\pm=-\beta k/(2\pi w_2^\mp)$, and this implies a simple modification of \eqref{rer}
\EQ{
w_1^+\,=\frac{\beta k}{2\pi}e^{2\pi\eta(t)/\beta}\ ,\qquad w_1^-\,=-\frac{\beta k}{2\pi}e^{-2\pi t/\beta}\,,\qquad\beta k\ll 1\,.
\label{pos}
}
This is the position of the QES represented by the blue blob in the AdS in figure \ref{fig7}.

Taking into account the modification of the conformal factor \eqref{xee} and the critical values \eqref{pos} yielding the position of the QES, gives the entropy
\EQ{
S_\text{QES in front}=2S_\text{BH}^{(\beta)}+\frac {\pi c}{3\beta}(t-\eta(t))-\frac c6\log\eta'(t)+2S_\text{shock}\ .
\label{grr}
}
Note that this equals $2S_\text{BH}^{(\beta)}+2S_\text{shock}$ at $t=t_0$, i.e.~the result \eqref{cat} with a jump by the entropy of the shockwave. 

As $t$ increases, the entropy starts to increase. After the time \eqref{tur}, $\eta(t)$ saturates to the constant value \eqref{his}. Past this time scale,  we have \eqref{bid}  and the entropy continues to increase linearly
\EQ{
S_\text{QES in front}\simeq 2S_\text{BH}^{(\beta)}+ \frac{\pi c}{3\beta}(t-\eta_0)+\frac{\pi c}{3\tilde\beta}(t-t_0)+\frac c6\log\frac{\beta^2-\tilde\beta^2}{4\beta^2}+2S_\text{shock}\ .
\label{yie}
}
At later times when the black hole is evaporating towards the temperature $\beta^{-1}$, we have \eqref{bom}, and the gradient of the entropy changes:
\EQ{
S_\text{QES in front}&\simeq 2S_\text{BH}^{(\beta)}+\frac{2\pi c }{3\beta}t+\frac{\pi c}{3\beta}(\kappa-t_0-\eta_0) +\frac c6\log\frac{\beta^2-\tilde\beta^2}{4\beta\tilde\beta}+2S_\text{shock}\ .
\label{yif}
}
Comparing \eqref{xie} and \eqref{yie} along with \eqref{xif} and \eqref{yif}, we note that, once the function $\eta(t)$ saturates, the difference  $\left(S_\text{no island}-S_\text{QES in front}\right)$ is independent of time.

\subsection{Island with QES behind the  shockwave}\label{s4.3}
 
For late times, indicated by the red region on the boundary, we expect the blue QES in figure \ref{fig7} to jump behind the shockwave, and become the red QES. When the QES is behind the shockwave it is more convenient to use a mixture of coordinates, 
\EQ{
w^+=e^{2\pi y^+/\beta}\ ,\qquad \tilde w^-=-\frac1{\hat f(y^-)}\ ,
}
as the vacuum coordinates. The choice is guided by the dilaton \eqref{los} which has a particularly nice expression in terms of the coordinates $(y^+,\tilde w^-)$. The boundary point then has the coordinates
\EQ{
y^+_2=t\ ,\qquad \tilde w_2^-=-\frac1{\hat f(t)}\ .
}
The conformal factors of the QES and the boundary point are\footnote{The conformal factor in $\tilde w^\pm$ coordinates is $\Omega^{-2}=4/(1+\tilde w^+\tilde w^-)^2$}
\EQ{
\Omega_1^{-2}=\frac{4}{(1+\tilde w_1^-\hat f(y_1^+))^2}\cdot \frac{\beta\hat f'(y_1^+)}{2\pi e^{2\pi y_1^+/\beta}}\ ,\qquad\qquad
\Omega_2^{-2}=\frac{\beta\hat f(t)^2}{2\pi e^{2\pi t/\beta}\hat f'(t)}\ .
}
Now we can write the generalized entropy, as a function of 
$(y_1^+,\tilde w_1^-)$ as 
\EQ{
S(y^+_1,\tilde w_1^-)=\frac{\phi_0}{2G_N}+\frac c3 F(y^+_1,\tilde w_1^-)+\frac c6\log \frac{\hat f(t)^2}{e^{2\pi t/\beta}\hat f'(t)}+\frac{c}3\log\frac{\beta}{\pi}\ .
}
Note that the region $D$ between the boundary and the QES no longer owns the shockwave and so there is no jump from the shockwave entropy. In the above, the function
\EQ{
F(y_1^+,\tilde w_1^-)&=\frac1{k}\left[\frac{\hat f''(y^+_1)}{2\hat f'(y^+_1)}-\frac{\tilde w_1^-\hat f'(y^+_1)}{1+\tilde w_1^-\hat f(y^+_1)}\right]
\\[5pt] &+\log\frac{\big(-e^{2\pi y_1^+/\beta}+e^{2\pi t/\beta}\big)\big(\tilde w_1^-+\hat f(t)^{-1}\big)}{1+\tilde w_1^-\hat f(y^+_1)}
+\frac12\log \frac{\hat f'(y_1^+)}{e^{2\pi y_1^+/\beta}}\ .
}

Now we extremize over the position of the QES $(y_1^+,\tilde w^-_1)$.
Extremizing over $\tilde w_1^-$ gives a linear equation for $\tilde w_1^-$ that can be solved
\EQ{
\tilde w_1^-=\frac{\hat f'(y_1^+)+k\hat f(y_1^+)-k\hat f(t)}{\hat f(t)(k\hat f(y_1^+)-\hat f'(y_1^+))-k\hat f(y_1^+)^2}\ .
\label{goo}
}
Then extremizing over $y_1^+$ gives a complicated equation  that can be solved numerically to determine $y_1^+$.  In general terms, the solution for 
the null coordinate $y_1^+$ of the QES lags behind the boundary time $t$ by an amount that we identify in the next section as the scrambling time:
\EQ{
\Delta t_\text{s}\,\equiv\,t-y_1^+.
}

In the early time regime, just after the shockwave enters the AdS region $t_0$,
it is more convenient to transform from $y_1^+$ to $\tilde w_1^+=\hat f(y_1^+)$. In the first instance, we are interested in the QES immediately after the shockwave enters, in which case we can use the approximation \eqref{bid} for $\hat f$, keeping the leading and next-to-leading terms,
\EQ{
y_1^+=\frac{\tilde\beta}{2\pi}\log\tilde w_1^++\frac{k\tilde\beta\xi}{4\pi}\log^2(e^{-2\pi t_0/\tilde\beta}\tilde w_1^+)+\cdots\ .
}
In the small $\beta k$ limit, the relevant terms are
\EQ{
F(\tilde w_1^\pm)&=\frac\pi{\tilde\beta k}\cdot\frac{1-\tilde w_1^+\tilde w_1^-}{1+\tilde w_1^+\tilde w_1^-}\big(1-k\xi\log(e^{-2\pi t_0/\tilde\beta} \tilde w_1^+)\big)
\\ &+\log\frac{(-(\tilde w_1^+)^{\tilde\beta/\beta}+e^{2\pi t/\beta})(\tilde w_1^-+e^{-2\pi t/\tilde\beta})}{1+\tilde w_1^+\tilde w_1^-}+\frac{\beta-\tilde\beta}{2\beta}\log \tilde w_1^++\cdots\ ,
}
where $\xi=(\beta^2-\tilde\beta^2)\tilde\beta/(4\pi\beta^2)$ comes from the next-to-leading term in \eqref{bid}. Extremizing over $\tilde w_1^\pm$ and keeping only the most dominant terms in the limit $\beta k\ll1$, gives\footnote{The key observation is that $\tilde w_1^+\tilde w_1^-$ is order $k$.}
\EQ{
\frac{2\pi}{\tilde\beta k}\tilde w_1^+-\frac1{\tilde w_1^-+e^{-2\pi t/\tilde\beta}}+\cdots&=0\ ,\\
\frac{2\pi}{\tilde\beta k}\tilde w_1^-+\frac{\pi\xi}{\tilde\beta \tilde w_1^+}-\frac{\beta-\tilde\beta}{2\beta \tilde w_1^+}+\cdots&=0\ .
}
These can be solved to yield the short-time behaviour
\EQ{
\tilde w_1^+=\frac{k\tilde\beta(3\beta-\tilde\beta)(\beta+\tilde\beta)}{8\pi\beta^2}e^{2\pi t/\tilde\beta}\ ,\qquad
\tilde w_1^-=\frac{(\beta-\tilde\beta)^2}{(3\beta-\tilde\beta)(\beta+\tilde\beta)}e^{-2\pi t/\tilde\beta}\ ,
\label{boo}
}
The solution for $\tilde w_1^+$ allows us to to extract the scrambling time
\EQ{
\Delta t_\text{s}=\frac{\tilde\beta}{2\pi}\log\frac{8\pi\beta^2}{k\tilde\beta(3\beta-\tilde\beta)(\beta+\tilde\beta)}\ .
\label{koo}
}

From \eqref{boo}, since $\tilde w_1^->0$, it follows that the QES is generically inside the horizon in this early time regime.

In the late time regime, $\hat f=e^{2\pi(t+\kappa)/\beta}$, the scrambling time satisfies
\EQ{
\sinh\frac{2\pi\Delta t_\text{s}}\beta=\frac{\pi}{\beta k}\ ,
}
with a solution
\EQ{
\Delta t_\text{s}=\frac{\beta}{2\pi}\log\frac{2\pi}{\beta k}\ ,
\label{noo}
}
at small $\beta k$. Then the solution of \eqref{goo} is
\EQ{
\tilde w_1^-=-\frac{\beta k}{2\pi}e^{-2\pi (t+\kappa)/\beta}\ .
\label{out}
}
So at late time, the QES is outside the horizon, as for an eternal black hole.

\pgfdeclareimage[interpolate=true,width=7cm]{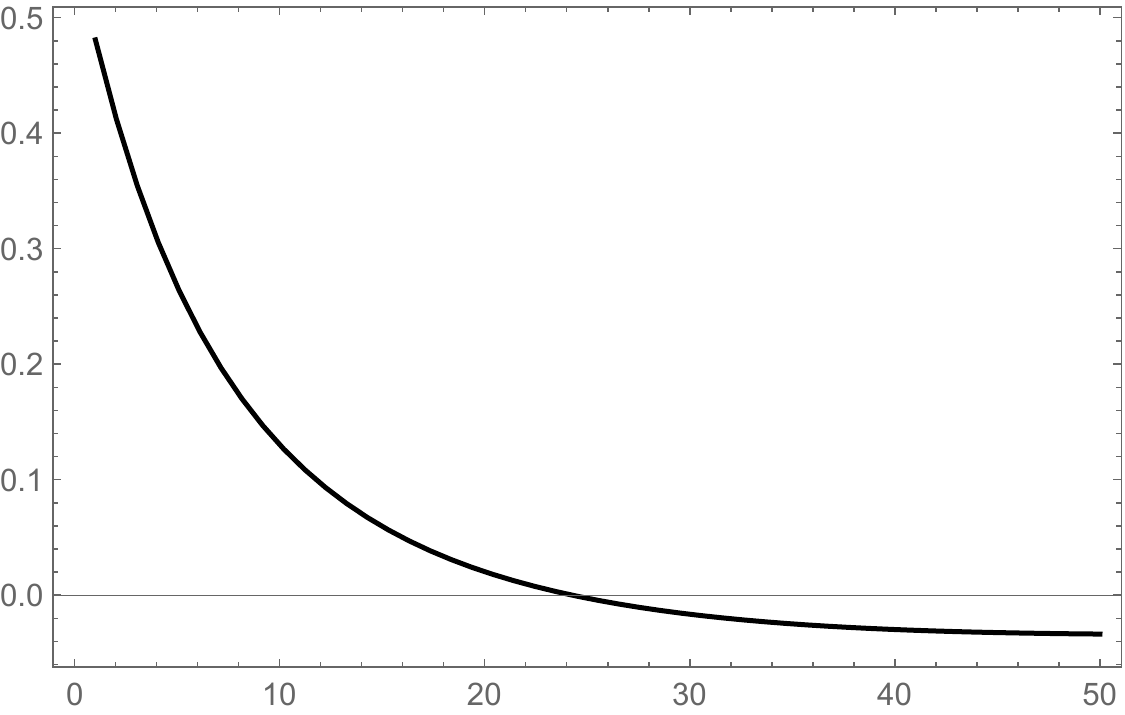}{fig3}
\pgfdeclareimage[interpolate=true,width=7cm]{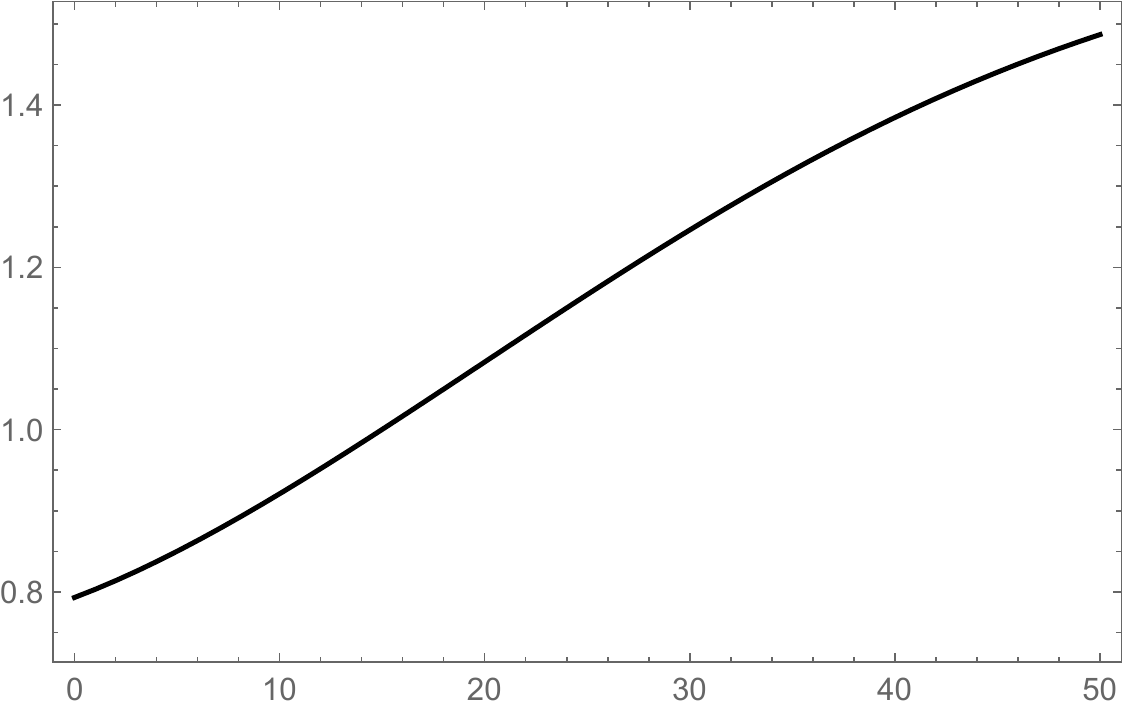}{fig7}
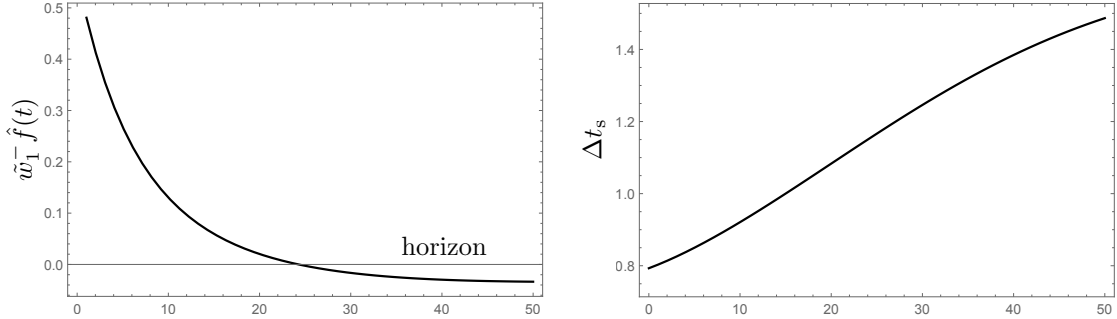
\begin{figure}
\begin{center}
\begin{tikzpicture}[scale=0.95]
\pgftext[at=\pgfpoint{0cm}{0cm},left,base]{\pgfuseimage{fig3}} 
\node at (5.6,1) {\footnotesize horizon};
\node[rotate=90] at (-0.3,2.5) {\footnotesize $\tilde w^-_1\hat f(t)$};
\node[rotate=90] at (7.7,2.5) {\footnotesize $\Delta t_\text{s}$};
\pgftext[at=\pgfpoint{8cm}{0cm},left,base]{\pgfuseimage{fig7}} 
\end{tikzpicture}
\caption{\footnotesize {\it Left\/}: a plot of $\tilde w_1^-\hat f(t)$ as a function of time, i.e.~the scaled position of the QES relative to the horizon, for some indicative values of the parameters, $\beta=3$, $\tilde\beta=1$ and $k=0.065$. Negative values correspond to points outside the horizon.  {\it Right\/}: the scrambling time. Horizon crossing occurs at time $\sim24$ with the origin of the time axis set at $t_0$. For the values of the parameters the scrambling time is $\sim0.8$. At short times \eqref{boo}, we have $\tilde w_1^-\hat f(t)\sim 0.5$ while at long times, the plot manifests the saturation in \eqref{out} $\beta k/2\pi\sim0.03$.}
\label{fig9}
\end{center}
\end{figure}

It is important that because of the fact that $y_1^+$ lags behind $t$, this entropy saddle can only appear for $t$ such that $y_1^+>t_0$. We can use the early time approximation \eqref{koo} to derive the condition in terms of the boundary time
\EQ{
t>t_0+\frac{\tilde\beta}{2\pi}\log\frac{8\pi\beta^2}{k\tilde\beta(3\beta-\tilde\beta)(\beta+\tilde\beta)}\ .
\label{kac}
}
So the saddle appears a scrambling time after the shockwave.

In the limits  we are working, the critical entropy is dominated by the values of the dilaton at the horizon, i.e.~the time-dependent expression $S_\text{BH}(y^+)$ in \eqref{due} where $y_1^+(t)$ is solution of the extremization problem:
\EQ{
S_\text{QES behind}=S_\text{BH}(y^+_1(t))+S_\text{cor.}\ .
\label{ser}
}
The correction $S_\text{corr.}$ remains subleading so the dominant contribution to the entropy of this saddle. Consequently, the entropy starts at $S_\text{BH}^{(\tilde\beta)}$ and then relaxes back to $S_\text{BH}^{(\beta)}$.

\subsection{Left/right independence}\label{s4.4}

In this section, we argue that our procedure of ignoring the effect of the left region of the two-sided black hole, on the right region is a valid one for times that are relevant to the competition of saddles around the Page time, i.e.~of order $k^{-1}$. The contribution to the entropy that we have ignored is the cross term
\EQ{
S_\text{cross}=\frac c6\log\frac{\sigma_{13}\sigma_{24}}{\sigma_{14}\sigma_{23}}\ ,
\label{xop}
}
where $\sigma_{ij}=-(w_i^+-w_j^-)(w_i^--w_j^-)$ is the spacetime interval.

We chose to have shockwaves symmetrically on both sides in order to simplify the discussion. The symmetry means that the QES and boundary points on the left are related to those on the right by $w_3^\pm=w_1^\mp$ and $w_4^\pm=w_2^\mp$, respectively. 

Let us consider the case with the QES in front of the shockwave since this is the case that is most sensitive to left-right effects because the two QES are closer. The equations of motion for the QES $p_1$ including the effects of the QES $p_3$ and boundary point $p_4$ on the left, are
\EQ{
\frac{2\pi}{\beta k}w^\pm_1=\frac1{w^\mp_1-w^\mp_2}+\frac1{w_1^\mp-w_3^\mp}-\frac1{w_1^\mp-w_4^\pm}\ .
}
We use the left-right symmetry to write these as 
\EQ{
\frac{2\pi}{\beta k}w^\pm_1=\frac1{w^\mp_1-w^\mp_2}+\frac1{w_1^\mp-w_1^\pm}-\frac1{w_1^\mp-w_2^\pm}\ .
\label{hex}
}
The question is whether the latter two terms on the right-hand side alter the solution we wrote down in \eqref{pos}. The coordinates $w_2^\pm$ are given in \eqref{cyr}. Now at relevant time scales, $t$ and $\eta(t)$ are order $k^{-1}$ and in the limit of small $\beta k$ it follows straightforwardly that the latter two terms in \eqref{hex} are either sub-leading, for the $w^+_1$ equation, or cancel, for the $w^-_1$ equation, and so do not change the solution \eqref{pos} at leading order. It is also simple to show that the cross terms \eqref{xop} are sub-leading to our expression for the entropy in \eqref{grr}.

The same reasoning and conclusion applies to the case when the QES are behind the shockwave.

\section{Page curves}\label{s5}

In this section, we find the Page curves for the shockwave scenario in JT gravity. We also discuss the associated scrambling time.

\subsection{Entanglement dynamics}

The Page curves are determined by finding the entropy saddles and then at any given boundary time, taking the one with the lowest entropy. This leads to transitions as the entropy of the saddles cross. These transitions point to fundamental re-arrangements of the entanglement structure of the black hole.

Before shockwave insertion, the eternal black hole in equilibrium with the radiation has an entropy transition at the late Page time in \eqref{fup}. Now we can consider what happens when we insert the shockwave. 
It is worth noting here that the shockwave carries energy and entropy like a large ``diary", to use the popular terminology \cite{Hayden:2007cs}. However, from the entropy point-of-view, the shockwave is not an analogue of a diary because its purifier is the radiation bath. On the contrary, the diary is assumed to be entangled with some auxiliary system. So shockwave entropy hastens the Page time because it increases the entanglement between the black hole and the radiation. The shockwave also carries energy, like the diary, that heats the black hole up and this has the opposite effect of delaying the Page time.

The issue of when entropy transitions occur, depends also on island contributions and several scenarios are possible. The entropy saddle corresponding to the red QES behind the shockwave, has the decreasing entropy \eqref{ser}. Note this saddle is delayed after the insertion by the scrambling time scale \eqref{kac}. So ultimately this will be the dominant saddle, but exactly what happens depends on all the parameters. Some potential scenarios are:

\pgfdeclareimage[interpolate=true,width=7cm]{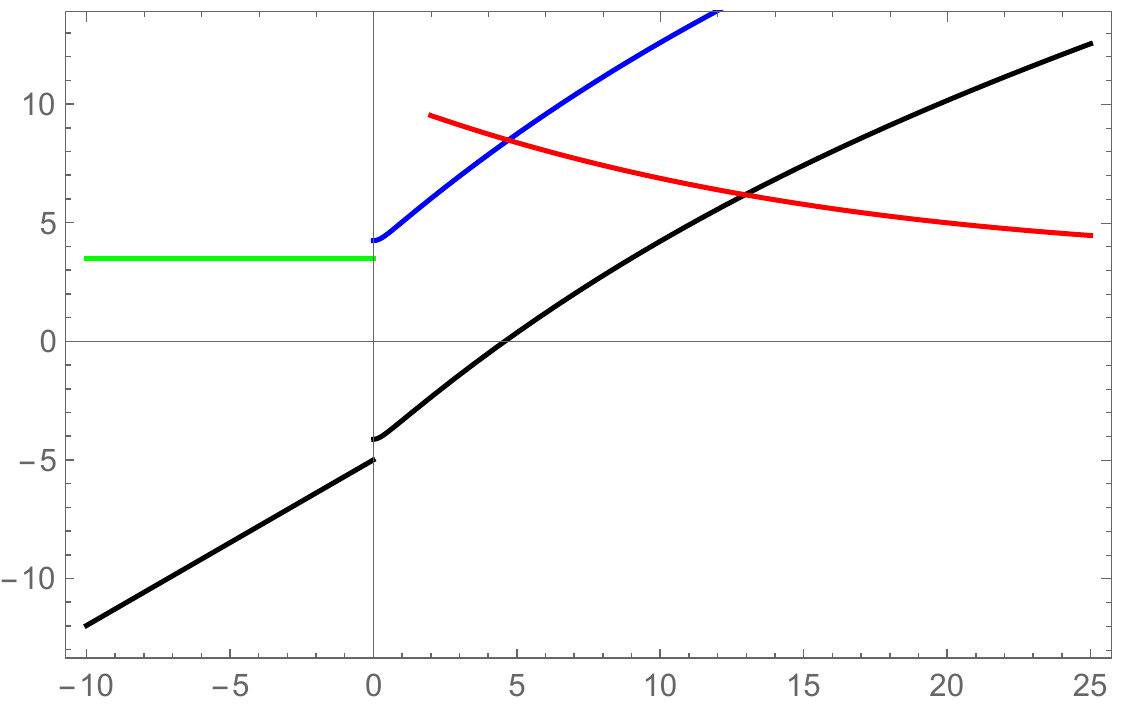}{fig1}
\pgfdeclareimage[interpolate=true,width=7cm]{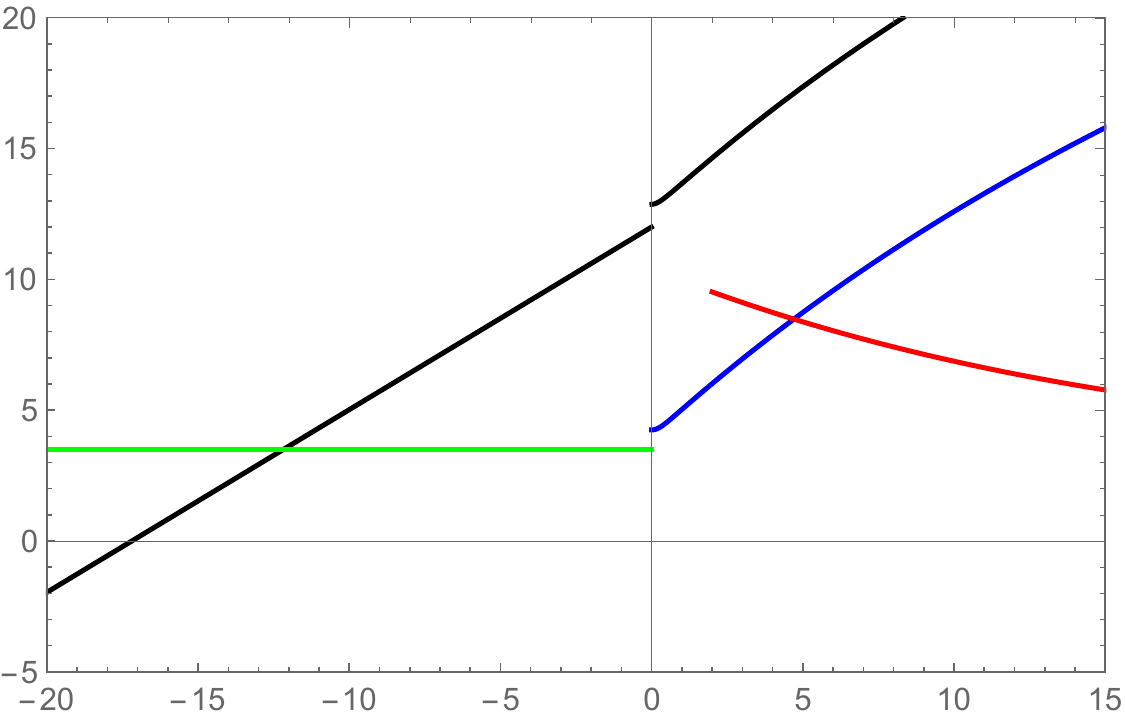}{fig2}
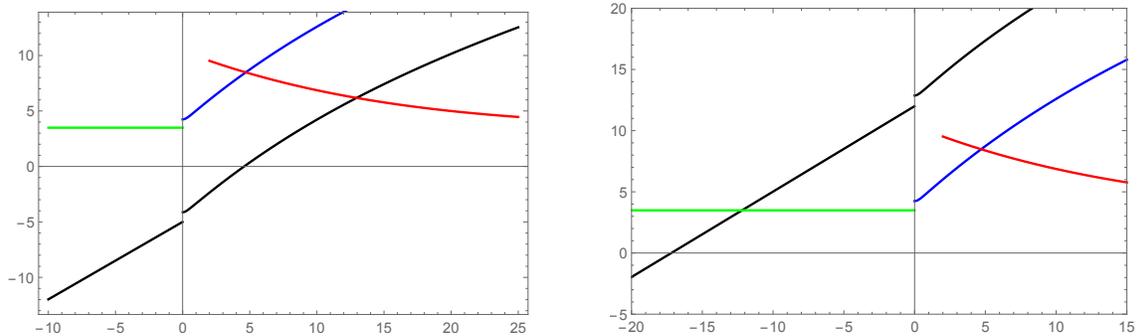
\begin{figure}
\begin{center}
\begin{tikzpicture}[scale=1]
\pgftext[at=\pgfpoint{0cm}{0cm},left,base]{\pgfuseimage{fig1}} 
\pgftext[at=\pgfpoint{8cm}{0cm},left,base]{\pgfuseimage{fig2}} 
\end{tikzpicture}
\caption{\footnotesize Two possible scenarios for the entropy saddles as a function of time discussed in the text with a small shockwave entropy. The origin of time is set at shockwave insertion (the vertical scale has also been shifted). Non-island saddles are shown in black and saddles with islands are coloured coded according to figure \ref{fig7} (green: boundary point in front of shockwave; blue: boundary point behind shockwave, QES in front; red: boundary point and QES behind shockwave). {\it Left\/}: insertion before the original Page time. {\it Right\/}: insertion after the original Page time. Note that the red saddle only comes into being a certain scrambling time after the insertion.}
\label{fig8}
\end{center}
\end{figure}

\noindent (i) The shockwave is inserted before the original Page time and the shockwave entropy is small (left side of figure \ref{fig8}). In this case, the original transition is avoided and a new Page time occurs when the no-island saddle jumps to the saddle with the QES  behind the shockwave.

\noindent (ii) The shockwave is inserted after the original Page time  and the shockwave entropy is small (right side of figure \ref{fig8}). After the insertion, the entropy of this saddle jumps and increases (in blue) until a new Page time is reached and eventually the red saddle dominates.

\noindent (iii) When the shockwave entropy is large, either of the scenarios in (i) and (ii) can lead to a delayed transition to the final saddle due to the time lag \eqref{kac}.

\subsection{Scrambling time}
 
\begin{figure}[ht]
\begin{center}
\begin{tikzpicture} [scale=0.8]
\filldraw[green!15] (-3,3.5) -- (0,0.5) -- (0,6) -- (-0.5,6) -- (-3,3.5); 
\filldraw[red!15] (-3,3.5) -- (-5.5,6) -- (-6,6) -- (-6,0.5) -- (-3,3.5); 
\draw[-] (0,0) -- (0,6);
\draw[very thick,dashed,->] (0,0.5) -- (-4,4.5);
\filldraw[red] (-3,3.5) circle (4pt);
\filldraw[red] (0,4.5) circle (4pt);
\draw [decorate,decoration={brace,amplitude=10pt}] (0.2,4.5) -- (0.2,0.5);
\node at (1.3,2.5) {\footnotesize $\Delta t_\text{s}$};
\node at (-2.5,6.5) (a1) {\footnotesize entanglement wedge};
\node at (-3.5,5.5) (a2) {\footnotesize QES};
\node at (-7.5,5) (a3) {\footnotesize island};
\node at (2.3,4.7) (a4) {\footnotesize boundary point};
\draw[->] (a1) -- (-1.5,4.5);
\draw[->] (a2) -- (-3.1,3.7);
\draw[->] (a3) -- (-5,4);
\draw[->] (a4) -- (0.2,4.6);
\end{tikzpicture}
\caption{\footnotesize The Hayden-Preskill protocol involves sending in a massless quantum into an old black hole (i.e.~one with a QES) from the boundary at some time $t'$. The quantum leaves the entanglement wedge of the boundary (green) and enters the island (pink) precisely when the quantum passes through the QES. This fixes the boundary time $t$ of the QES. The difference $t-t'$ is the scrambling time of the black hole $\Delta t_\text{s}$.}
\label{fig4} 
\end{center}
\end{figure}
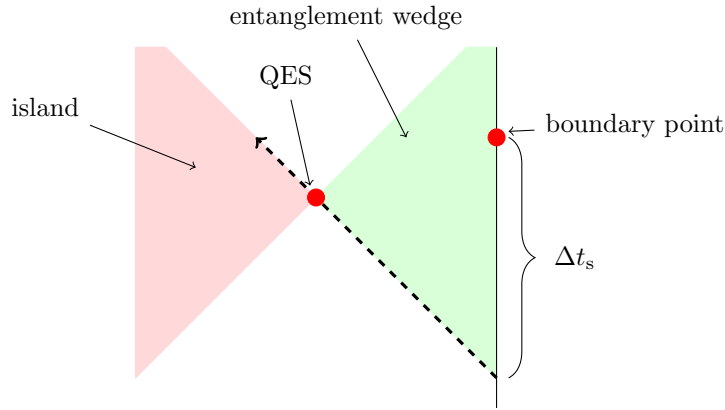

The Hayden-Preskill protocol interprets the scrambling time as the minimum time it takes for quantum information thrown into an old black hole to be recoverable in the Hawking radiation \cite{Hayden:2007cs}. In the present context, to be recoverable from the Hawking radiation is interpreted as being in the island. So the scrambling time is the difference of the current boundary time with the boundary time of an in-going null ray that just passes through the QES: see figure \ref{fig4}. The implication is that for a wavepacket sent in more than a scrambling time in the past will be in the island, rather than the entanglement wedge of the boundary and hence be recoverable from the full radiation Hilbert space.

\paragraph{QES in front of shock:}
Before the shockwave, the scrambling time takes the value \eqref{scr}. As the boundary point passes the shockwave at $t=t_0$, the QES has coordinates \eqref{pos}.
Since $\eta(t)$ increases  as $t$, the QES starts on the same Cauchy slice as the boundary point, but then starts to lag behind the Cauchy slice as $\eta(t)$ saturates. After this time the QES moves towards the horizon in the $w^-$ direction but $w^+$ becomes frozen at the value
\EQ{
w^+_\text{frozen}=\frac{\beta k}{2\pi}\cdot\frac{\beta+\pi f(\infty)}{\beta-\pi f(\infty)}=\frac{\beta k}{2\pi}\cdot\frac{\beta+\tilde\beta}{\beta-\tilde\beta}e^{2\pi t_0/\beta}\ .
}
If the saturation happens quickly, i.e.~when for large shockwave energy $\tilde\beta\ll\beta$, then $\eta(t)\approx t_0$ and the boundary time corresponding to the null coordinate $w^+$ of the QES is frozen at a scrambling time before the time when the shockwave goes in. This is in agreement with the observations of Penington \cite{Penington:2019npb} concerning the generic effect of large diaries on the QES of black holes after the Page time.

During this regime the scrambling time is effectively time dependent. A
wave packet sent in at boundary time $t'$ leaves the entanglement wedge of the boundary at a time $t$, where
\EQ{
e^{2\pi t'/\beta}=\frac{\beta k}{2\pi}e^{2\pi\eta(t)/\beta}\ ,
}
and so the effective, time dependent, scrambling time is
\EQ{
\Delta t_\text{s}=t-t'=\frac\beta{2\pi}\log\frac{2\pi}{\beta k}+t-\eta(t)\ .
}
When $\eta(t)$ saturates, the scrambling time increases linearly as
\EQ{
\Delta t_\text{s}=t-t_0+\frac\beta{2\pi}\log\frac{2\pi}{\beta k}-\frac{\beta}{2\pi}\log\frac{\beta+\tilde\beta}{\beta-\tilde\beta}\ .
}

\paragraph{QES behind shock:}
Finally, the red saddle point, for which the QES is behind the shockwave, the scrambling time is determined by the solution for $\Delta t_\text{s}=t-y_1^+$ of the extremization problem. In the early time regime, $\hat f(t)\sim e^{2\pi t/\tilde\beta}$, this is precisely \eqref{koo}:
\EQ{
\Delta t_\text{s}=\frac{\tilde\beta}{2\pi}\log\frac{8\pi\beta^2}{k\tilde\beta(3\beta-\tilde\beta)(\beta+\tilde\beta)}\ .
}
In the late time regime, $\hat f(t)=e^{2\pi(t+\kappa)/\beta}$ and the scrambling time returns to that of the original black hole \eqref{scr}. The scrambling time for this saddle is shown in figure \ref{fig9}.

\subsection{QES: inside or outside the horizon?}
 
An interesting issue is whether the QES of the final saddle that dominates the entropy, i.e.~the red one in figure \ref{fig7}, is inside or outside the new horizon. 

The $\tilde w_1^-$ coordinate in the early time regime was found in \eqref{boo}. So, generically, the QES is inside the horizon unless the shockwave energy is very small given that $\beta k\ll1$.

In the long time regime, the QES is found outside the horizon \eqref{out}, as one expects for an eternal black hole. The conclusion that we draw from this is that non-equilibrium conditions maintained for a long time, i.e.~large shockwave energy and slow evaporation, favour the QES to be inside the horizon.

\section{Shockwaves and the extremal black hole}\label{s6}

In this section, we consider the same shockwave set up but where the initial black hole is extremal. This leads to a scenario has some similarity to that considered in \cite{Almheiri:2019psf} and the solution for the function $f(t)$ will be the same.

\subsection{The extremal black hole}

For this case, the radiation bath has zero temperature and the black hole is one sided. The extremal black hole corresponds to a solution with $f(t)=t$.
The dilaton takes the form
\EQ{
\phi=\phi_0+\frac{2\phi_r}{x^--x^+}\ .
} 

What is perhaps surprising, is that the extremal black hole is at an entropy saddle with an island \cite{Almheiri:2019yqk}.\footnote{On the other hand there is no left-hand side for the cut of a no-island saddle to end on.} In order to find it, we write the generalized entropy with the QES with coordinates $x_1^\pm$. As previously, for simplicity we take the point in the bath to be just at the boundary with Schwarzschild time $t$,
\EQ{
S_\text{gen.}(x_1^\pm)=\frac{\phi_0}{4G_N}+\frac c6\left(\frac1{k}\cdot\frac{1}{x_1^--x_1^+}+\log\frac{2(t-x_1^+)(x_1^--t)}{x_1^--x_1^+}\right)\ .
\label{luk}
} 
Extremizing over the position of the QES, for small $k$, gives
\EQ{
x_1^\pm=t\mp \frac1{2k}
}
This implies an island  extending a small way outside the horizon. The critical entropy  is the Bekenstein-Hawking entropy plus a small correction:
\EQ{
S_\text{gen.}=\frac{\phi_0}{4G_N}+\frac c6-\frac c6\log(2k)\ .
}
The scrambling time is large:
\EQ{
\Delta t_\text{s}=t-x_1^+=\frac1{2k}\ .
}

\subsection{The solution with a shockwave}

We choose to insert the shockwave at $t=0$ with an energy that defines the temperature $\beta^{-1}$:
\EQ{
E_\text{shock}=\frac{\pi\phi_r}{4G_N\beta^2}\ .
}
Hence, the analogue of \eqref{lop} is
\EQ{
\{f(t),t\}=-2\pi^2\beta^{-2}e^{-kt}\ ,
\label{lop2}
} 
which can be solved in terms of Bessel functions by 
\EQ{
\hat f(t)=\frac{e^{4\pi/(\beta k)}}\pi\frac{K_0(z)}{I_0(z)}\ ,\qquad z=\frac{2\pi}{\beta k}e^{-kt/2}\ .
\label{huf}
}
This defines a particular solution, precisely the one in \cite{Almheiri:2019psf}. The pre-factor has been chosen for later convenience.
The solution for $f(t)$ is then, as before, a M\"obius transformation 
\EQ{
f(t)=\frac{A\hat f(t)+B}{C\hat f(t)+D}\ ,
\label{fee}
}
fixed by requiring the initial conditions dictated by the extremal black hole, $f(0)=f''(0)=0$ and $f'(0)=1$. This determines
\EQ{
&A=\beta e^{-4\pi/(\beta k)}\, I_0\big(\tfrac{2\pi}{\beta k}\big)\ ,\quad B=-\frac\beta\pi\, K_0\big(\tfrac{2\pi}{\beta k}\big)\ ,\\[5pt]
&C= \pi e^{-4\pi/(\beta k)}\,I_1\big(\tfrac{2\pi}{\beta k}\big)\ ,\quad D=K_1\big(\tfrac{2\pi}{\beta k}\big)\ .
}
In the early time window $t\ll k^{-1}$, taking into account the pre-factor in 
\eqref{huf} we have,
\EQ{
\hat f(t)\,\simeq\, e^{2\pi t/\beta}\ .
\label{dee}
}
There is a longer window $t\ll k^{-1}|\log\beta k|$, since we work with $\beta k\ll1$, for which we have the approximation
\EQ{
\hat f(t)\thicksim \exp\left[\frac{4\pi}{\beta k}\left(1-e^{-kt/2}\right)\right]\ .
\label{gre}
}
Finally, in the long time regime $t\gg k^{-1}|\log\beta k|$, we have
\EQ{
\hat f(t)\thicksim \frac{e^{4\pi/(\beta k)}}\pi\left(\frac{kt}2-\gamma+\log\frac{\beta k}{\pi}\right)\ .
\label{vax}
}
Since this is linear in $t$, it manifests a return to the extremal solution and so at long times  the excited black hole settles back to the extremal one.

The Bekenstein-Hawking entropy defined \eqref{due} is
\EQ{
S_\text{BH}(y^+)=\frac1{4G_N}\left(\phi_0+\phi_r\cdot\frac{\hat f''(y^+)}{\hat f'(y^+)}\right)
=\frac{1}{4G_N}\left(\phi_0+\phi_r\cdot\frac{kz I_1(z)}{K_0(z)}\right)\ .
\label{cir}
}
where $z$ is the function defined in \eqref{huf} with $t$ replaced by $y^+$.

\pgfdeclareimage[interpolate=true,width=7cm]{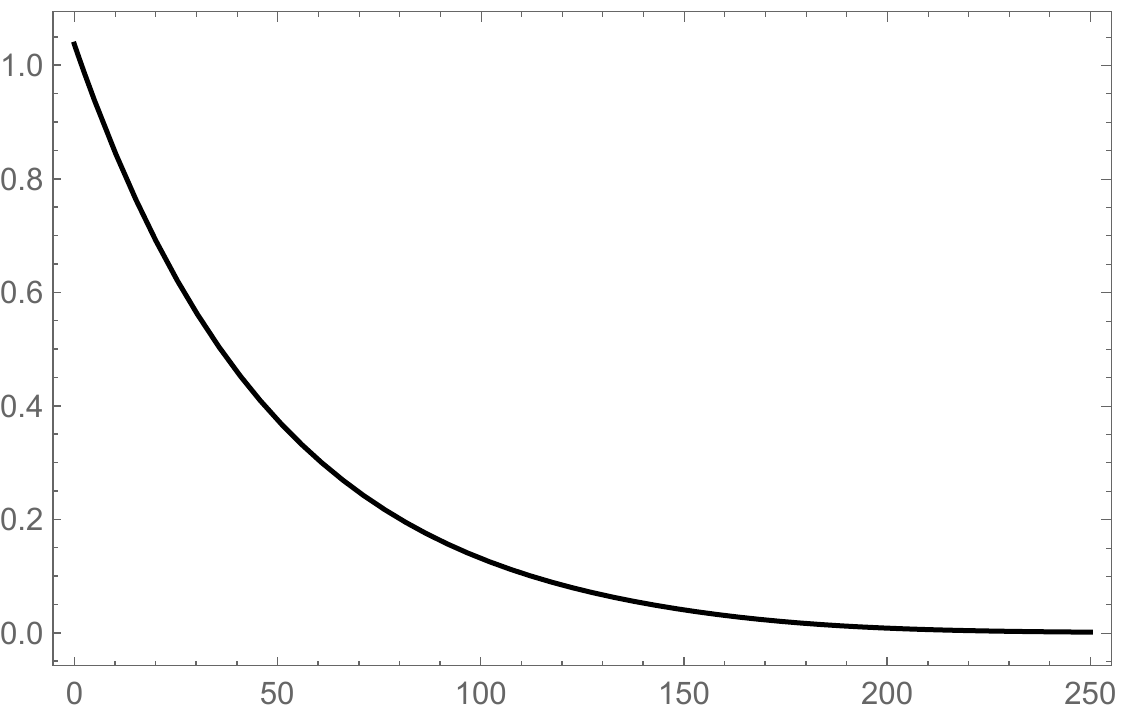}{fig8}
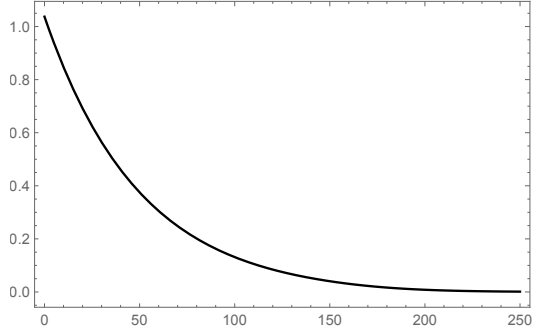
\begin{figure}
\begin{center}
\begin{tikzpicture}[scale=1]
\pgftext[at=\pgfpoint{0cm}{0cm},left,base]{\pgfuseimage{fig8}} 
\end{tikzpicture}
\caption{\footnotesize The Bekenstein-Hawking entropy of the evaporating black hole after shockwave insertion. The vertical axis has been scaled appropriately.}
\label{fig10}
\end{center}
\end{figure}

In calculating the entropy, it is useful to notice that the stress tensor component for the outgoing modes  $T_{x^-x^-}=0$ whilst for the ingoing modes it is $T_{y^+y^+}=0$. So $y^+$ and $x^-$, or any coordinate related to these by a M\"obius transformation, are  vacuum coordinates.

\subsection{QES in front of shockwave}

Let us calculate the position of the QES and the entropy, when the QES is in front of the shockwave. The boundary point is behind the shockwave, so has coordinates
\EQ{
y_2^+=t\ ,\qquad x^-_2=f(t)\ .
}
On the other hand, the QES is in front of the shockwave, and so we can use $x_1^+=y_1^+$ and $x_1^-$ as coordinates. The conformal factors of the QES and the boundary are
\EQ{
\Omega_1^{-2}=\frac{4}{(x^-_1-x^+_1)^2}\ ,\qquad\Omega_2^{-2}=\frac1{f'(t)}\ .
}
Using these we can compute the generalized entropy,
\EQ{
S_\text{gen.}(x_1^\pm)=\frac{\phi_0}{4G_N}+\frac c6\left(\frac1{k}\cdot\frac{1}{x_1^--x_1^+}+\log\frac{2(t-x_1^+)(x_1^--f(t-b))}{(x_1^--x_1^+)\sqrt{f'(t)}}\right)+S_\text{shock}\ .
\label{luk}
}

For small $k$, and assuming that $kt\ll1$, the QES is at
\EQ{
x_1^+=\frac{3f(t)-t}2-\frac1{2k}\ ,\qquad x_1^-=\frac{3t-f(t)}2+\frac1{2k}\ ,
}
and so the scrambling time is,
\EQ{
\Delta t_\text{s}=t-x_1^+=\frac1{2k}+\frac32\big(t-f(t)\big)\ .
}

As $t$ increases and becomes of $\mathscr{O}(k^{-1})$, $x_1^\pm$ remain  $\mathscr{O}(k^{-1})$, and so in the small $k$ limit, with $kt$ and $\beta$ fixed, we can use the approximate form \eqref{gre} to find the leading order behaviour of the entropy at $\mathscr{O}(k^{-1})$. The behaviour at this order is driven by the conformal factor of the point in the bath:
\EQ{
S_\text{QES in front}
=\frac{\phi_0}{4G_N}+\frac{\pi c}{3\beta k}\big(1-e^{-kt/2}\big)+S_\text{shock}+\cdots\ ,
\label{sck}
}
where the corrections involve $\log k$. This growing entropy is shown on the left of figure \ref{fig9} in black.

\subsection{QES behind shockwave}

The second entropy saddle has the QES behind the shockwave. In this case, it is more convenient to use coordinates $w^\pm$ related to $x^\pm$ by 
\EQ{
x^+=\frac{Aw^++B}{C w^++D}\ ,\qquad x^-=\frac{A-Bw^-}{C-Dw^-}\ .
}
so that
\EQ{
w^+=\hat f(y^+)\ ,\qquad w^-=-1/\hat f(y^-)\ .
}
The horizon is at $x^-=f(\infty)=A/C$ behind the shockwave, and this corresponds to $w^-=0$, so this coordinate is a good choice when the QES can be both inside or outside the horizon.

The vacuum coordinates we will use are $y^+$ and $w^-$. The boundary point is at
\EQ{
y_2^+=t\ ,\qquad w^-_2=-1/\hat f(t)\ .
}
The conformal factors of the QES and boundary point are
\EQ{
\Omega_1^{-2}=\frac{4\hat f'(y_1^+)}{(1+w_1^-\hat f(y_1^+))^2}\ ,\qquad\Omega_2^{-2}=\frac{\hat f(t)^2}{\hat f'(t)}\ .
}

Now we can write the generalized entropy, as a function of $(y_1^+,w_1^-)$ as 
\EQ{
S_\text{gen.}(y^+_1,w_1^-)=\frac{\phi_0}{4G_N}+\frac c6 F(y^+_1,w_1^-)+\frac c{12}\log \frac{4\hat f(t)^2}{\hat f'(t)}\ .
}
The region $D$ between the boundary and the QES does not now host the shockwave and there is no jump from the shockwave entropy. The function $F$ to be extremized is,
\EQ{
F(y_1^+,w_1^-)&=\frac1{k}\left\{\frac{\hat f''(y_1^+)}{2\hat f'(y_1^+)}-\frac{w_1^-\hat f'(y_1^+)}{1+w_1^-\hat f(y^+_1)}\right\}\\[5pt]
&+
\log\frac{(t-y_1^+)(w_1^-+1/\hat f(t))}{1+w_1^-\hat f(y_1^+)}+\frac12\log \hat f'(y_1^+)\ .
}
We can follow the same approach as in section \ref{s4.3} to find the QES in the regime just after the shockwave enters the AdS region at $t_0$. First of all, we change variable from $y_1^+$ back to $w_1^+=\hat f(t)$, in the early time regime, we have
\EQ{
y_1^+=\frac\beta{2\pi}\log w_1^++\frac{k\beta^2}{16\pi^2}\log^2w_1^++\cdots\ .
}
In the small $k$ limit, the relevant terms are
\EQ{
F(w_1^\pm)&=\frac\pi{\tilde\beta k}\cdot\frac{1- w_1^+w_1^-}{1+w_1^+w_1^-}\Big(1-\frac{k\beta}{4\pi}\log w_1^+\Big)
\\ &+\log\frac{(t-\frac\beta{2\pi}\log w_1^+)(w_1^-+1/\hat f(t))}{1+w_1^+w_1^-}+\frac12\log w_1^++\cdots\ ,
}
Extremizing over $w_1^\pm$ and keeping only the most dominant terms in the limit $\beta k\ll1$, gives\footnote{The key observation is that $w_1^+w_1^-$ is order $k$.}
\EQ{
\frac{2\pi}{\beta k}w_1^+-\frac1{w_1^-+1/\hat f(t)}+\cdots&=0\ ,\\
\frac{2\pi}{\beta k}w_1^--\frac1{4w_1^+}+\cdots&=0\ .
}
These can be solved to yield the coordinates of the QES in the early time regime after the shockwave enters
\EQ{
w_1^+=\frac{3\beta k}{8\pi}\hat f(t)\ ,\qquad
w_1^-=\frac1{3\hat f(t)}\ .
\label{boo2}
}
In the short-time limit, $\hat f(t)\approx\exp(2\pi t/\beta)$ and we can extract the scrambling time 
\EQ{
\Delta t_\text{s}=t-y_1^+=\frac\beta{2\pi}\log\frac{8\pi}{3\beta k}\ .
}
It is apparent that the QES is behind the horizon at early times. In addition, this saddle only appears when $t>\Delta t_\text{s}$.

\pgfdeclareimage[interpolate=true,width=7cm]{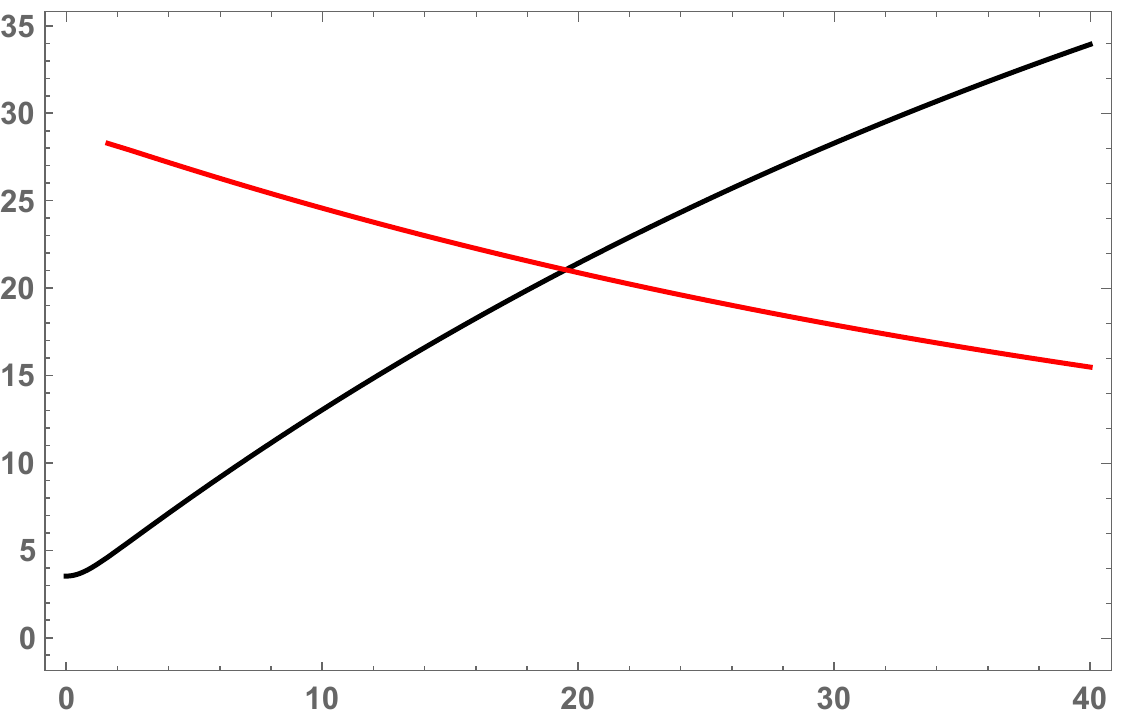}{fig5}
\pgfdeclareimage[interpolate=true,width=7cm]{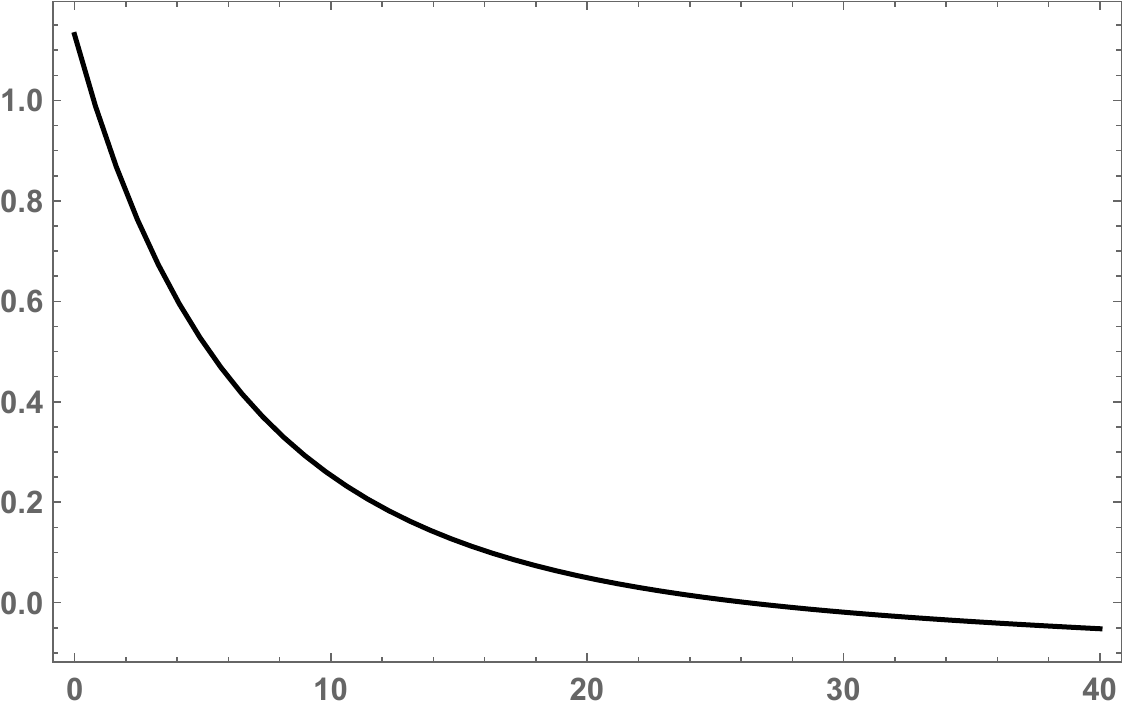}{fig6}
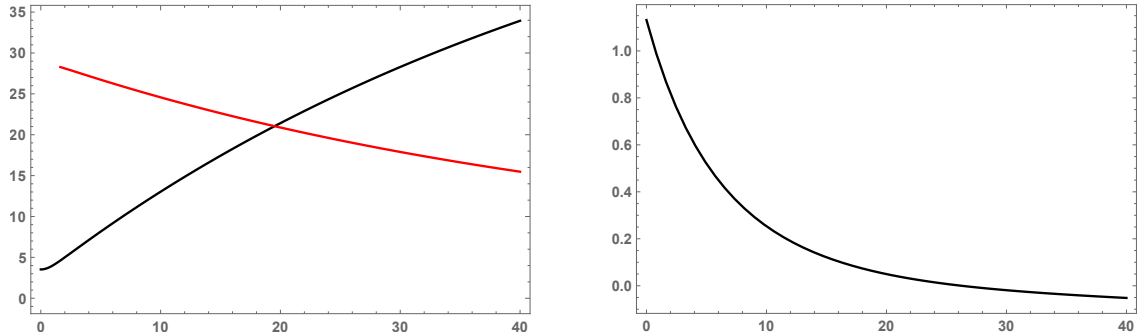
\begin{figure}
\begin{center}
\begin{tikzpicture}[scale=1]
\pgftext[at=\pgfpoint{0cm}{0cm},left,base]{\pgfuseimage{fig5}}
\pgftext[at=\pgfpoint{8cm}{0cm},left,base]{\pgfuseimage{fig6}}
\end{tikzpicture}
\caption{\footnotesize {\it Left\/}: the Page curve, with the saddle with the QES in front/behind the shockwave in black/red. {\it Right\/}: the $w_1^-$ coordinate of the QES as a function of time. The QES starts behind the horizon but moves outside at a later time, around $t=30$ here.}
\label{fig9}
\end{center}
\end{figure}

At early times $t\ll k^{-1}$, the QES lies behind the horizon and so the critical entropy is close to  the Bekenstein-Hawking entropy of the evaporating black hole in \eqref{cir} with $t$ replaced by the null coordinate of the QES $y_1^+=t-\Delta t_\text{s}$. At later times, with $kt$ fixed and $k$ small, we can use the approximation \eqref{gre} to find the leading order behaviour of the entropy at order $k^{-1}$,
\EQ{
S_\text{QES behind}
=\frac{\phi_0}{4G_N}+\frac{\pi c}{6\beta k}e^{-kt/2}+\cdots\ .
}
This decaying entropy is shown on the left of figure \ref{fig9} in red. By equating this to \eqref{sck}, and assuming that the shockwave entropy is small, we can extract the Page time
\EQ{
t_\text{Page}=\frac2k\log\frac32\ ,
} 
in the small $k$ limit.

At much later times, when $\hat f(t)$ is approximated by \eqref{vax}, i.e.~the form for an extremal black hole, the QES is outside the horizon. So just as in the finite temperature case, the QES begins inside the horizon but then moves outside as equilibrium is restored.

\section{Conclusions}\label{s7}

We have analysed the way the entanglement structure of the finite temperature and extremal black holes in JT gravity is modified when a CFT shockwave is inserted into them. The back-reaction problem can be solved exactly and then the entropy saddles can be found by using the generalized entropy prescription. More fundamentally, we expect that the latter would follow from replica wormholes in the presence of the shockwave and we leave the demonstration of this to future work. 

The shockwaves carry energy and entropy into the black hole that affects the entanglement structure in quite complicated ways that depend on the parameters. The entanglement re-arrangement at the Page time is generally disrupted. The Page time can be hastened or postponed and there can be additional Page times as the QES jumps from being in front of the shockwave to being behind. 

Another interesting phenomenon, is the behaviour of QES relative to the horizon of the black hole. In equilibrium, the QES is generally outside the horizon \cite{Almheiri:2019yqk} but when the shockwave is inserted, the equilibrium is disturbed, the black hole starts to evaporate and the QES is inside the horizon. As evaporation proceeds and the black hole returns to equilibrium, the QES moves from the inside to the outside.

\acknowledgments We would like to thank Justin David, Carlos N\'u\~nez, Dan Thompson and Andrea Legramandi for discussions. TJH and SPK acknowledge support from STFC grant ST/P00055X/1.

\appendix
\appendixpage

\section{Coordinate systems}
\label{appcoords}

In our analysis, we use various coordinate systems, each of which has its utility, depending on process or time scale  of interest. Here we collect together these different coordinate systems for easy reference.

The Poincar\'e patch of AdS$_2$ is covered by $(x^+, x^-)$ coordinates:
\EQ{
ds^2=-\frac{4dx^+dx^-}{(x^+-x^-)^2}\ .
}
The $(y^+, y^-)$ coordinates cover the Schwarzschild black hole patch:
\EQ{
x^\pm=\frac{\beta}{\pi}\tanh\frac{\pi y^\pm}{\beta}\,,\qquad ds^2=-\frac{4\pi^2}{\beta^2}\frac{dy^+ dy^-}{\sinh^2\frac\pi\beta(y^+-y^-)}\ .
}
In {\em front of the shockwave}, the coordinates $(w^+, w^-)$   are related to $x^\pm$ and $y^\pm$ (for the black hole on right side) as,
\EQ{
w^\pm&=\pm e^{\pm 2 \pi y^\pm/\beta}\,,\qquad  x^\pm=\pm\frac\beta\pi\cdot\frac{w^\pm \mp1}{w^\pm\pm1}\,,\\[5pt]
ds^2&=-\frac{4dw^+ dw^-}{(1+w^+ w^-)^2}\,.
}
For points {\em behind the shockwave}, the relation between the Poincar\'e patch coordinates and  $y^\pm$ changes:
\begin{eqnarray}
x^\pm=f(y^\pm)=\pm\frac\beta\pi\cdot\frac{w^\pm \mp1}{w^\pm\pm1}\,.
\end{eqnarray}
where $f(y^\pm)$ is fixed by the M\"obius transformation in terms of $\hat f(y^\pm)$ in eq.\eqref{exactf}. 
The coordinates $\left(\tilde x^+, \tilde x^-\right)$  and $(\tilde w^+, \tilde w^-)$
are also naturally used behind the shockwave with
\begin{eqnarray}
\tilde x^\pm=\pm\frac{\tilde\beta}{\pi}\cdot\frac{\tilde w ^\pm \mp 1}{\tilde w^\pm \pm 1}\,,
\end{eqnarray}
and 
\begin{eqnarray}
\tilde w^\pm=\pm \hat f(y^\pm)^{\pm 1}\,.
\end{eqnarray}

\section{Exact solution for $f(t)$} 

The exact solution to the differential equation \eqref{lop} is in terms of modified Bessel functions, with a particular solution (choosing $\alpha=1$)
\EQ{
\hat f (t)=\frac{K_\nu(\nu z)}{I_\nu(\nu z)}\,,\quad \nu=\frac{2\pi}{\beta k}\,.
}
The specific solution $f(t)$ which satisfies the boundary conditions \eqref{pop} is a M\"obius transform of $\hat f(t)$
\EQ{
f(t)=\frac{A \hat f +B}{C \hat f +D}\ .
}
The constants $\{A,B,C,D\}$ can be  fixed up to an overall (irrelevant) multiplicative constant by the matching conditions \eqref{pop}. We find,
\EQ{
A&=\aleph
I_\nu(\nu z_0)\,
\left[1\,+\,z_0\, \frac{I_\nu^\prime(\nu z_0)}{I_\nu(\nu z_0)}\tanh\left(\frac{\pi t_0}{\beta}\right)\right]\ ,\\[5pt]
B&=- \aleph
K_\nu(\nu z_0)\,
\left[1\,+\,z_0\,\frac{K_\nu^\prime(\nu z_0)}{K_\nu(\nu z_0)}\tanh\left(\frac{\pi t_0}{\beta}\right)\right]\ ,\\[5pt]
C&=\aleph\frac{\pi}{\beta}
\,I_\nu(\nu z_0)\,
\left[\tanh\left(\frac{\pi t_0}{\beta}\right)\,+\,z_0\, \frac{I_\nu^\prime(\nu z_0)}{I_\nu(\nu z_0)}\right]\ ,\\[5pt]
D&=-\aleph\frac{\pi}{\beta}
\,K_\nu(\nu z_0)\,
\left[\tanh\left(\frac{\pi t_0}{\beta}\right)\,+\,z_0\, \frac{K_\nu^\prime(\nu z_0)}{K_\nu(\nu z_0)}\right]\ ,
}
where, the (irrelevant) constant $\aleph$ can be chosen to set 
$AD-BC=1$:
\EQ{
\aleph=\sqrt{\frac{k}{2}}\,\cosh\left(\frac{\pi t_0}{\beta}\right)\,,\qquad\implies\ AD - BC=1\,.
}

\section{High temperature limit}\label{sec:appadia}

When $\beta k\ll 1$, the index $\nu$ of the Bessel functions is large and we can then use the integral representions for $I_\nu$ and $K_\nu$ to deduce a saddle point (or WKB-like) expression for the function $f(t)$. Consider the integral representation of the modified Bessel function of the first kind
(for ${\rm Re}\, \nu>-\tfrac12$)
\EQ{
I_\nu(\nu z)=\left(\frac{\nu z}{2}\right)^\nu\frac{1}{\sqrt\pi\Gamma(\nu+\tfrac12)}\int
_0^\pi e^{\nu z \cos\theta}\left(\sin\theta\right)^{2\nu}\,d\theta\,.
}
In the large $\nu$ limit the integral is dominated by a saddle point and evaluating the leading contribution from the saddle point, we get,
\EQ{
I_\nu(\nu z) \big|_{\nu \gg 1}\simeq \frac{1}{\sqrt\pi}\,\exp\left[\nu\left(\sqrt{1+z^2}-{\rm tanh}^{-1}\frac{1}{\sqrt{1+z^2}}\right)\right]\ .
}
Similarly, for the modified Bessel function of the second kind, we can make use of its integral representation, 
\begin{equation}
K_\nu(\nu z)=\sqrt{\pi}\left(\frac{\nu z}{2}\right)^\nu\frac{1}{\Gamma(\nu+\tfrac12)}\int
_0^\infty e^{-\nu z \cosh t}\left(\sinh t\right)^{2\nu}\,dt\,.
\end{equation}
Once again the large $\nu$ saddle point approximation can be employed to yield,
\EQ{
K_\nu(\nu z) \big|_{\nu \gg 1}\simeq \sqrt\pi\,\exp\left[-\nu\left(\sqrt{1+z^2}-{\rm tanh}^{-1}\frac{1}{\sqrt{1+z^2}}\right)\right]\,.
}
Therefore the function $\hat f (t)$ in \eqref{wup} can be given a WKB-like form in the adiabatic limit
\EQ{
\hat f(t)\big|_{\nu\gg 1}\approx\pi \alpha\exp\left[2\nu {\cal S}(t)\right]\ ,
\label{fadia}
}
where ${\cal S}(t)$ is defined in \eqref{jip}. 

We may also write ${\cal S}(t)$ in the WKB-like integral form
\EQ{
\nu\left({\cal S}(t)-{\cal S}(t_0)\right)=\pi\int_{t_0}^t\frac{dt^\prime}{\beta_{\rm eff}(t^\prime)}\, \ ,
\label{swkb}
}
where 
\EQ{
\beta_{\rm eff}^{-1}\,\equiv\, \beta^{-1}\sqrt{1+e^{-k(t-t_0)}E_\text{shock}/E_\beta}\,.
}

\end{document}